\documentclass[a4paper,twocolumn,nofootinbib,superscriptaddress]{revtex4-1}

\usepackage{xcolor}
\usepackage{hyperref}
\hypersetup{
	colorlinks,
	linkcolor={red!90!black},
	citecolor={red!10!blue},
	urlcolor={blue!80!black}
}

\usepackage{amssymb}
\usepackage{bm,amsmath}
\usepackage{graphicx}
\usepackage{array}

\newcommand{\p}{\partial}    
\newcommand{\f}{\frac}

\renewcommand{\d}{\mathrm{d}}

\renewcommand{\l}{\lambda}

\newcommand{\DD}{\Delta}

\newcommand{\mH}{\mathcal{H}}
\newcommand{\Teff}{T_{\text{eff}}}
\newcommand{\nhat}{\hat{\mathbf{n}}_i}

\usepackage{amssymb}
\usepackage{gensymb}

\begin{document}

\title{A perspective on active glassy dynamics in biological systems} 

\author{Souvik Sadhukhan}
\affiliation{Tata Institute of Fundamental Research, Hyderabad - 500046, India}

\author{Subhodeep Dey}
\affiliation{Tata Institute of Fundamental Research, Hyderabad - 500046, India}

\author{Smarajit Karmakar}
\email{smarajit@tifrh.res.in}
\affiliation{Tata Institute of Fundamental Research, Hyderabad - 500046, India}

\author{Saroj Kumar Nandi}
\email{saroj@tifrh.res.in}
\affiliation{Tata Institute of Fundamental Research, Hyderabad - 500046, India}

\begin{abstract}
Dynamics is central to living systems. In the last two decades, experiments have revealed that the dynamics in diverse biological systems - from intracellular cytoplasm to cellular and organismal aggregates - are remarkably similar to that in dense systems of inanimate particles in equilibrium. They show a glass transition from a solid-like jammed state to a fluid-like flowing state, where a moderate change in control parameter leads to an enormous variation in relaxation time. However, biological systems have crucial differences from the equilibrium systems: the former have activity that drives them out of equilibrium, novel control parameters, and enormous levels of complexity. These active systems showing glassy dynamics are known as active glasses. The field is at the interface of physics and biology, freely borrowing tools from both disciplines and promising novel, fascinating discoveries. We review the experiments that started this field, simulations that have been instrumental for insights, and theories that have helped unify diverse phenomena, reveal correlations, and make novel quantitative predictions. We discuss the primary characteristics that define a glassy system. For most concepts, we first discuss the known equilibrium scenario and then present the key aspects when activity is introduced. We end the article with a discussion of the challenges in the field and possible future directions. 
\end{abstract}

\maketitle

\section{Introduction}

\begin{figure*}
 	\includegraphics[width=17cm]{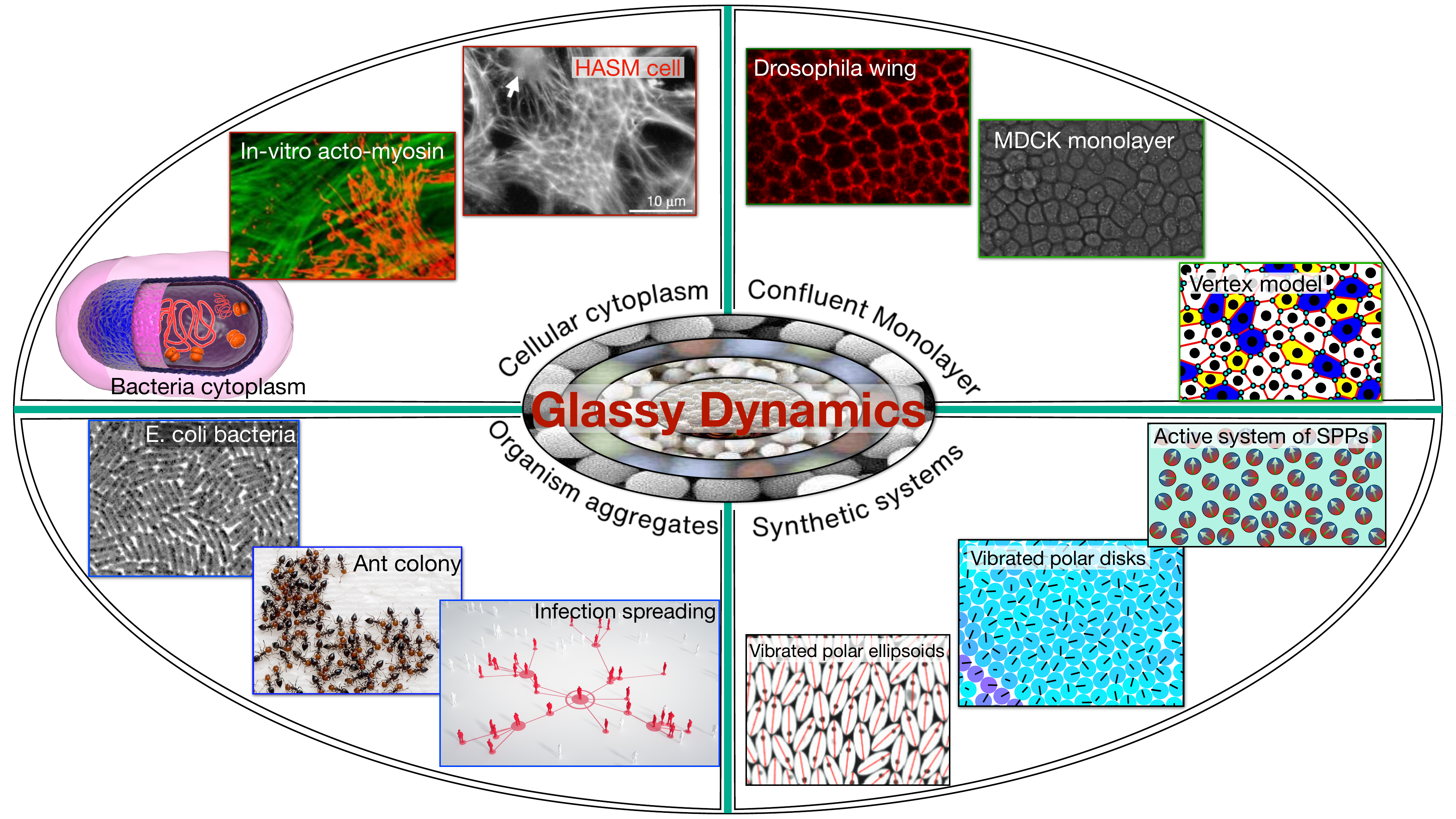}
 	\caption{We show different examples of biological and biology-inspired systems having glassy dynamics. We present four different classes of systems; despite their widely varying natures, they all have glass-like dynamics. Some of the figures are taken with permission from Refs.  \cite{lama2022,theers2018,fabry2001,Arora2022,Berthier2011a}. }
 	\label{examples_glass}
 \end{figure*}

This review concerns the fascinating phenomenology of glassy dynamics in biological systems at varying length scales. Glassy dynamics refers to the extreme dynamical slowdown, by several orders of magnitude, with a modest change in the control parameters \cite{Berthier2011a,gotzebook}. Surprisingly, the phenomenon does not accompany any phase transition or discernible change in static structures. A snapshot of a liquid and a glass look nearly identical, but their dynamics are markedly different. Glassy systems show slower than exponential (stretched-exponential) relaxation \cite{phillips1996}, sub-diffusive mean-square displacement (MSD) at intermediate times \cite{weeks2002}, non-Gaussian distribution of particle displacement \cite{Chaudhuri2007}, dynamical heterogeneity \cite{franz2000}, aging \cite{cugliandolo1993,Nandi2012}, etc. In the last couple of decades, experiments have shown that many biological systems also have glass-like dynamics. Examples include the cell cytoplasm \cite{fabry2001,bursac2005,Trepat2009,Nishizawa2017}, cellular aggregates and tissues \cite{Angelini2011,Park2015a,Garcia2015,Malinverno2017,Giavazzi2017,Mongera2018,Rode2019}, colonies of bacteria \cite{Takatori2020} or ants \cite{Gravish2015,Tennenbaum2016,Helbing2005}, synthetic systems \cite{Deseigne2010,Kumar2014,Klongvessa2019a}, etc. This glass transition from a solid-like jammed state to a fluid-like flowing state seems to be crucial for several biologically significant processes, such as wound healing \cite{poujade2007,das2015,Brugues2014,Malinverno2017}, cancer progression \cite{Streitberger2020,Friedl2003a}, embryogenesis \cite{Tambe2011,Friedl2009b,Malmi-Kakkada2018,Schotz2013}, and many others. The importance of the problem has led to many simulations \cite{Ni2013,Berthier2014b,Mandal2016b,Flenner2016,Bi2016,Bi2014,Berthier2017} and theories \cite{Berthier2013,Szamel2014a,Szamel2016,Szamel2015b,feng2017,liluashvili2017,Nandi2017b,Nandi2018a,Debets2022} for a quantitative understanding of the problem. Figure \ref{examples_glass} provides some examples of various biological systems having glassy dynamics that was once the subject of inert systems alone. These examples, and many others, have immensely enriched the field of glassy dynamics with new challenges, fresh ideas, and possibilities of novel discoveries.

One essential feature of biological systems is that they are active: the constituent particles consume energy and do some work. The work can be diverse: for example, the particles can divide, die, differentiate, be confluent, change their conformation, control the geometry and strength of interaction, propel themselves, etc \cite{jacques2015,samreview}. Developing a theoretical framework for such systems is a daunting task. However, commendable research works of the last decades have shown that it is possible to reveal the generic principles and obtain a theoretical framework for these systems, at least in some appropriate limits \cite{Marchetti2013,jacques2015,samreview}. Predictions made from such theories have been tested and validated in experiments and simulations. For example, cell division and apoptosis fluidize the system by cutting off the relaxation time scale \cite{Ranft2010,silke2017}. Different stochastic models can make robust predictions about cellular fate \cite{gilbertbook}. Energy landscape ideas of statistical physics provide crucial insights into the protein folding pathways and distinctive folding processes \cite{bryngelson1995}. These fascinating examples of applying physics principles to complex systems demonstrate that it is possible to draw meaningful insights via the consideration of specific aspects of these systems at a time. In the last decade or so, a large amount of theoretical work has focused on the glassy dynamics in active systems of self-propelled particles (SPPs) and confluent epithelial tissues. 
 
Active systems of SPPs comprise particles with a self-propulsion force, $f_0$, and a persistence time, $\tau_p$, of their motion \cite{Ramaswamy2010,Vicsek2012,Marchetti2013,Bechinger2016}. Many biological systems can be conveniently modelled as systems of SPPs; for example, birds and fishes \cite{Gueron1996,Cavagna2010,Pavlov2000,Mukundarajan2016}, ants colonies \cite{Gravish2015}, swimming bacteria \cite{Patteson2015}, etc. There are also examples from cellular \cite{Angelini2011,Garcia2015,Park2015a,Mongera2018,Rode2019} and sub-cellular levels \cite{Julicher2007,Schaller2010,Parry2014}, as well as synthetic systems \cite{Mijalkov2016, Scholz2018,Nsamela2022,Dauchot2005,Narayan2007,Kumar2014,Kelly1999,Palacci2010,Jiang2010}. Properties of active SPP systems in their dilute regime have been the subject of intense research activities in the last several decades \cite{vicsek1995,Marchetti2013,Ramaswamy2010}. It is well-known that these systems show many non-trivial properties. For example, they can have a flocking transition in spatial dimension two when the mean velocity or the average direction of the particles go from zero to a non-zero value \cite{vicsek1995}. We know that a continuous symmetry cannot spontaneously break in spatial dimension two; this is the Mermin-Wagner theorem \cite{kardarbook}. However, this theorem does not apply to active systems as they are out of equilibrium. 
In fact, a recent work \cite{galliano2023} has shown that it is not only the orientation order, nonequilibrium fluctuations in active systems can be strong enough to violate the Mermin-Wagner theorem leading to translational order as well.
Reference \cite{dey2024enhanced} has shown that long wavelength density fluctuations, reminiscent of Mermin-Wagner like fluctuations, in 2D active glasses with only a few percent of active particles performing run and tumble active motions get enhanced by several factors leading to divergence of mean squared position fluctuations with increasing system size $L$ in a power-law as $\langle\Delta r^2\rangle \sim L^{\delta}$, with $\delta \sim 1$ rather than usual $\log({L})$ divergence as predicted by Mermin-Wagner theorem in equilibrium solids.
Similarly, even the disordered phase is quite different from ordinary liquids. These systems show giant number fluctuations \cite{Narayan2007,Marchetti2013}: the particle number fluctuation, $\Delta N$, in a specific volume is proportional to the average number of particles $N$ in that volume. By contrast, ordinary liquids have $\Delta N\sim \sqrt{N}$. Another surprising aspect of active systems is the presence of long-range velocity correlations in these systems \cite{Marchetti2013,caprini2020,caprini2020b}. In contrast, dense active systems are subjects of more recent interest. Experiments reveal that they have glass-like properties \cite{fabry2001,Garcia2015,Angelini2011,Zhou2009}.

On the other hand, epithelial tissues have quite a distinctive character compared to ordinary particulate systems. Epithelial tissues are confluent, i.e., packing fraction remains unity at all times. This specific character enforces different types of models to theoretically study their properties. Some such models are the Vertex model \cite{Farhadifar2007,Fletcher2014,albert2016}, the Voronoi model \cite{Sussman2018b,Paoluzzi2021}, the cellular Potts model \cite{Graner1992,Glazier1993,Hogeweg2000}, etc. These models represent cells as polygons and can be considered either in \cite{Sadhukhan2021a,hirashima2017} or out of equilibrium, depending on the absence or presence of self-propulsion \cite{Bi2016,Paoluzzi2021}, cell division and apoptosis \cite{Ranft2010,Sadhukhan2022}. In recent literature, the competition between molecular crowding and thermal or active agitation leading to slow dynamics has sometimes been described as jamming \cite{Berthier2019c,Atia2021}. We emphasize that the term jamming here is different from the zero-temperature zero-activity geometric transition in disordered systems \cite{mari2009,berthier2009,biroli2013}. In this field, it refers to the transition separating the solid-like and the fluid-like states. This transition, strictly speaking, is the glass transition. However, as the term jamming is easy to grasp, it has grown in popularity \cite{Berthier2019c,Atia2021,Sadhukhan2022}.

Biological systems are complex, with too many variables. One must selectively choose the relevant parameters for the phenomenon of interest. Choosing the ``right model'', for example, particulate vs. confluent, as discussed above, is also crucial for theoretical progress. What are the benefits of theoretical analysis, particularly for such complex systems? It is often instrumental for deeper insight and quantitative predictions. But apart from these, it also ``{\it reveals relations between quantities or phenomena that would go unnoticed without a theoretical model}'' \cite{samreview}. 
On the other hand, from the physics perspective, these fascinating systems extend the scope and extent of the equilibrium glass transition problem. Many of these phenomena, exhibited in active glasses, are amenable to rigorous theoretical frameworks. Understanding these characteristics can lead to deeper insights into the equilibrium problem itself.

\begin{figure*}
	\includegraphics[width=14.6cm]{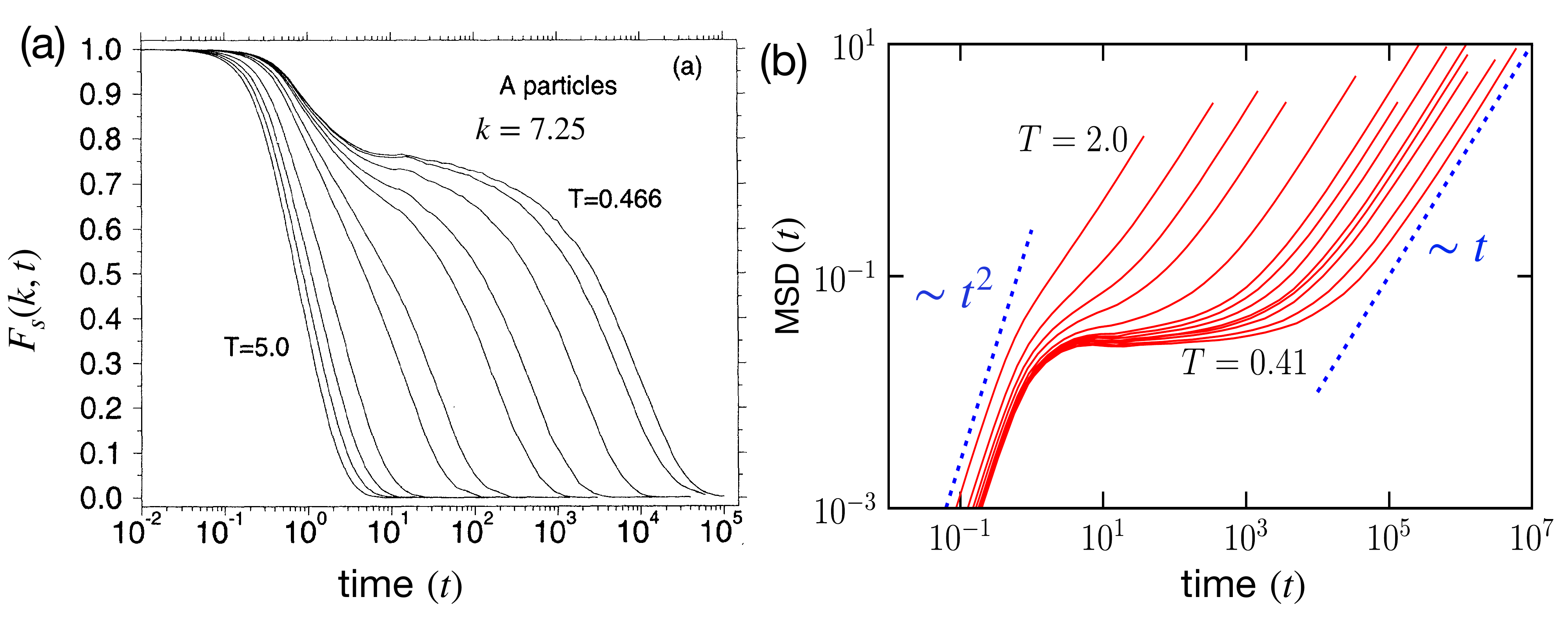}
	\caption{(a) The self intermediate scattering function, $F_s(k,t)$, at wave vector $k=7.25$ as a function of time $t$ for various $T$ (Adapted with permission from Ref. \cite{Kob1995b}). (b) Mean-square displacement (MSD) as a function of $t$ for different $T$. MSD changes ballistically at short times, goes to a sub-diffusive plateau at intermediate times, and becomes diffusive at long times. The plateau increases as $T$ decreases (Adapted with permission from Ref. \cite{Berthier2011a}).}
	\label{relaxationdyn}
\end{figure*}

The field of active glass is necessarily interdisciplinary. For the most part, we, therefore, take a parallel approach. We will first briefly discuss the known results of the equilibrium problem and then present the corresponding results for active glasses. We hope such a presentation will benefit an wider audience. We emphasize that the term ``active'' is quite broad and can refer to several forms, not limited to self-propulsion alone \cite{jacques2015,samreview}. 
There already exist several reviews summarizing various aspects of active glasses \cite{Berthier2019c,Janssen2019Review,Tailleur2022Book,Atia2021}. Our perspective article complements these excellent reviews. This review is organized as follows: we first describe the defining characteristics of a glassy system in Sec. \ref{glassdef}. We then briefly summarize in Sec. \ref{expts} some of the experimental results that led to this field, followed by a summary of simulations in Sec. \ref{simulations}. We review the theoretical developments in Sec. \ref{theory} and conclude this review in Sec. \ref{conclusion}, discussing the current status and our perspective on the future directions and challenges of the field.

\section{How to characterize a glassy system} 
\label{glassdef}

Glass transition refers to the change of the liquid-like state to the solid-like state without crystallization when we vary some system parameters, such as temperature or density. The relaxation time, $\tau$, and the viscosity, $\eta$, increase rapidly as the temperature $T$ decreases. The glass transition temperature, $T_g$, is the $T$ at which $\tau$ becomes a specific value, $\sim 10^2-10^3s$ (say). Here, we first discuss how to characterize a glassy system; these characteristics are the same for any system in the glassy regime. The most common defining hallmark of glassy systems is the slower than-exponential, i.e., stretched exponential relaxation \cite{Cavagna2009,Berthier2011a,Karmakar2014,vaibhav2020}. One can characterize this via the self-intermediate scattering function, $F_s(k,t)$, at wave vector $k$, and time $t$,
\begin{align}\label{fskt}
F_s(k,t)= \frac{1}{N}\Big \langle \sum_{i = 1}^{N} e^{\dot{\iota} \mathbf{k}.(\mathbf{r}_{i}(t) - \mathbf{r}_{i}(0))} \Big \rangle,
\end{align}
where $N$ is the number of particles, $\mathbf{r}_{i}(t)$ is the position of the $i$th particle at $t$, and $\langle \ldots\rangle$ denotes ensemble as well as time origin averaging. Another measure that is often used in the study of the dynamics of supercooled liquids is the overlap function, $Q(t)$ \cite{guiselin2020,Cavagna2009}, defined as:
\begin{equation}\label{Qoft}
Q(t)=\langle \tilde{Q}(t)\rangle= \Big \langle\frac{1}{N}\sum_{i=1}^{N} W\Big(a - |\mathbf{r}_{i}(t)-\mathbf{r}_{i}(0)|\Big)\Big \rangle,
\end{equation}
where $W(x)$ is the Heaviside Step Function: $W(x)$ is 1 if $x>0$ and 0 otherwise. The parameter $a$ represents the typical vibrational amplitude of the caged particles. $F_s(k,t)$ and $Q(t)$ show exponential decay in a liquid. Relaxation becomes complex close to $T_g$: they decay towards a plateau at intermediate times and then towards zero at long times (Fig. \ref{relaxationdyn}a) \cite{Cavagna2009,Berthier2011a,Karmakar2014}. The long-time data fit well with a stretched exponential form, $\phi(t)\sim \exp[-(t/\tau)^\beta]$, $\beta$ is the stretching exponent. When $F_s(k,t)$ or $Q(t)$ decays to a particular value, usually taken as $1/e$, that time defines a relaxation time, $\tau$.

 \begin{figure}
	\includegraphics[width=8.6cm]{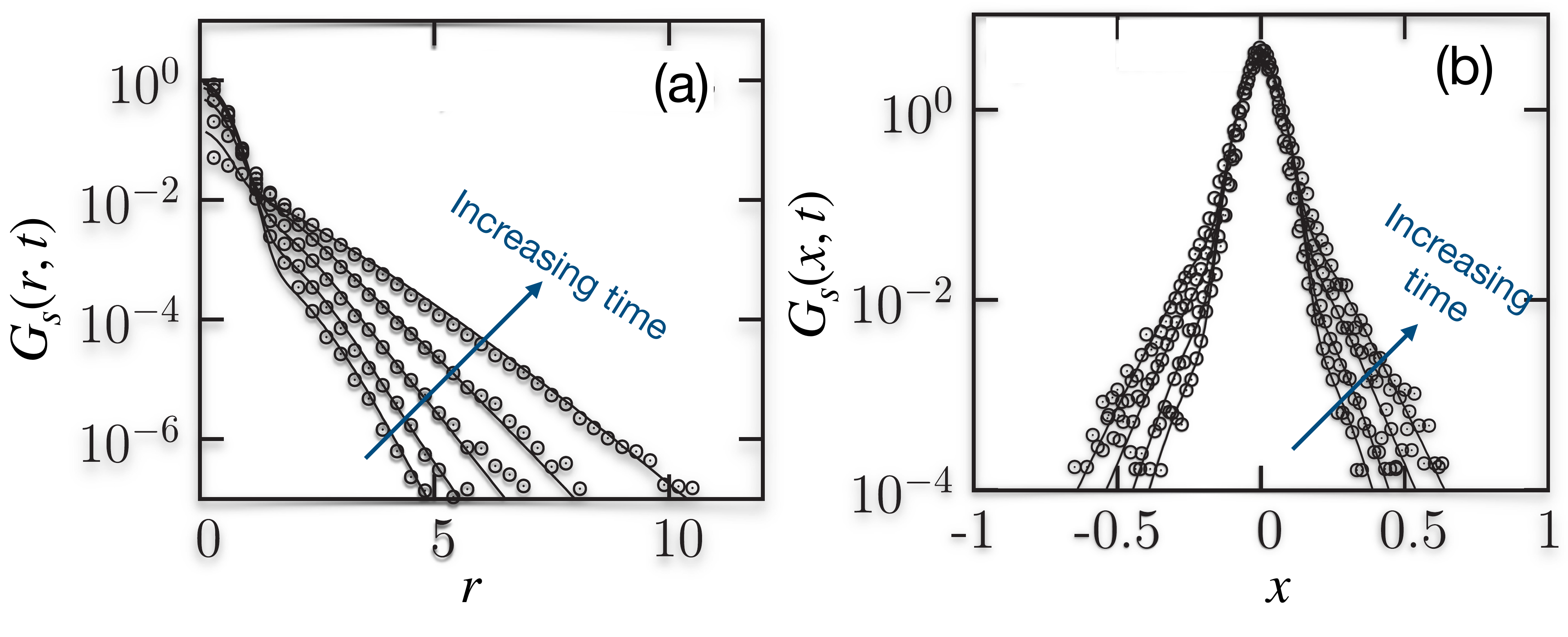}
	\caption{(a) The van-Hove function for spatial displacement, $G_s(r,t)$, and (b) the van-Hove function for displacement in a particular direction, $G_s(x,t)$. The directions of increasing time are shown in the figure. Adapted with permission from Ref. \cite{Chaudhuri2007}.}
	\label{vanHove}
\end{figure}

\begin{figure*}
	\includegraphics[width=15cm]{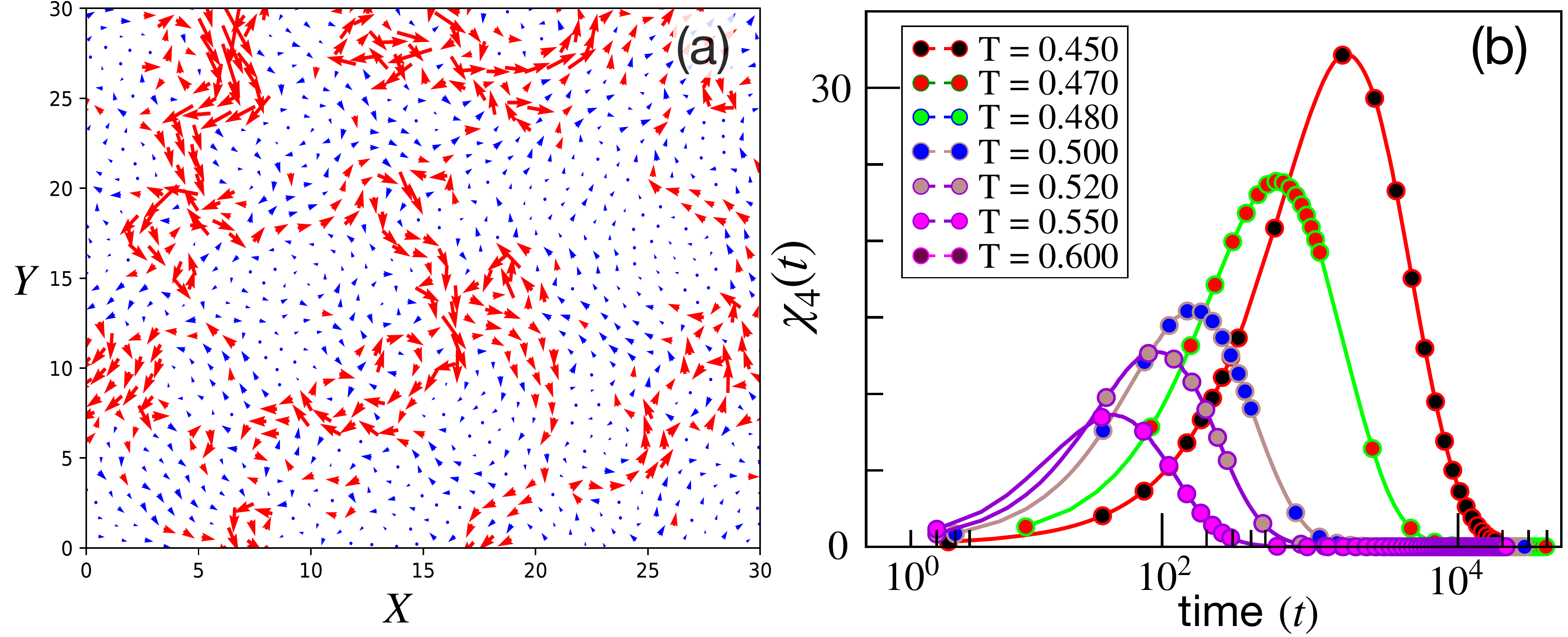}
	\caption{(a) Snapshot of a glassy system with the arrow length proportional to velocity and the colors red and blue for the fast and slow (compared to the average velocity) particles shows dynamical heterogeneity. (b) The four-point function, $\chi_4(t)$, characterizes DH; it has a non-monotonic nature, with the peak time corresponding to a relaxation time and the peak value to the domain volume. [Adapted with permission from Ref. \cite{smarajitPNAS2009}].}
	\label{DH_chi4}
\end{figure*}

When molecular crowding is dominant compared to thermal fluctuations, it is hard for a particle to move through the other particles as their movement is also constrained, leading to the phenomena of caging, another hallmark glassy characteristic. 
We can track the motion of an average particle via mean-square displacement (MSD) at time $t$:
\begin{equation}\label{MSDoft}
\text{MSD}(t)=\Big \langle\frac{1}{N} \sum_{i=1}^N [r_i(t)-r_i(0)]^2\Big \rangle.
\end{equation}
The particle moves freely up to a very short inter-particle distance, manifested by the ballistic part of MSD with slope $2$ (Fig. \ref{relaxationdyn}b). After that, it feels the presence of the other particles, and the movement gets constrained; it vibrates inside the cage formed by the neighboring particles. MSD becomes flat and sub-diffusive at this intermediate time. At very long times, it breaks the cage and gets trapped in another cage. This hopping-like motion is a universal feature of glassy relaxation, leading to a universal exponential tail in the van Hove correlation function (discussed later). Subsequent breakage of cages eventually leads to a diffusive motion at a long enough timescale. This transition from sub-diffusive to diffusive behavior is another generic feature of glassy systems. We emphasize that although glassy systems show sub-diffusive MSD and stretched exponential auto-correlation functions, these characteristics alone do not imply glassy dynamics. Several non-glassy systems can also show these characteristics \cite{vaibhav2020,mukherjee2024}. Glassy systems show several additional nontrivial features.

\begin{figure}
	\includegraphics[width=7.6cm]{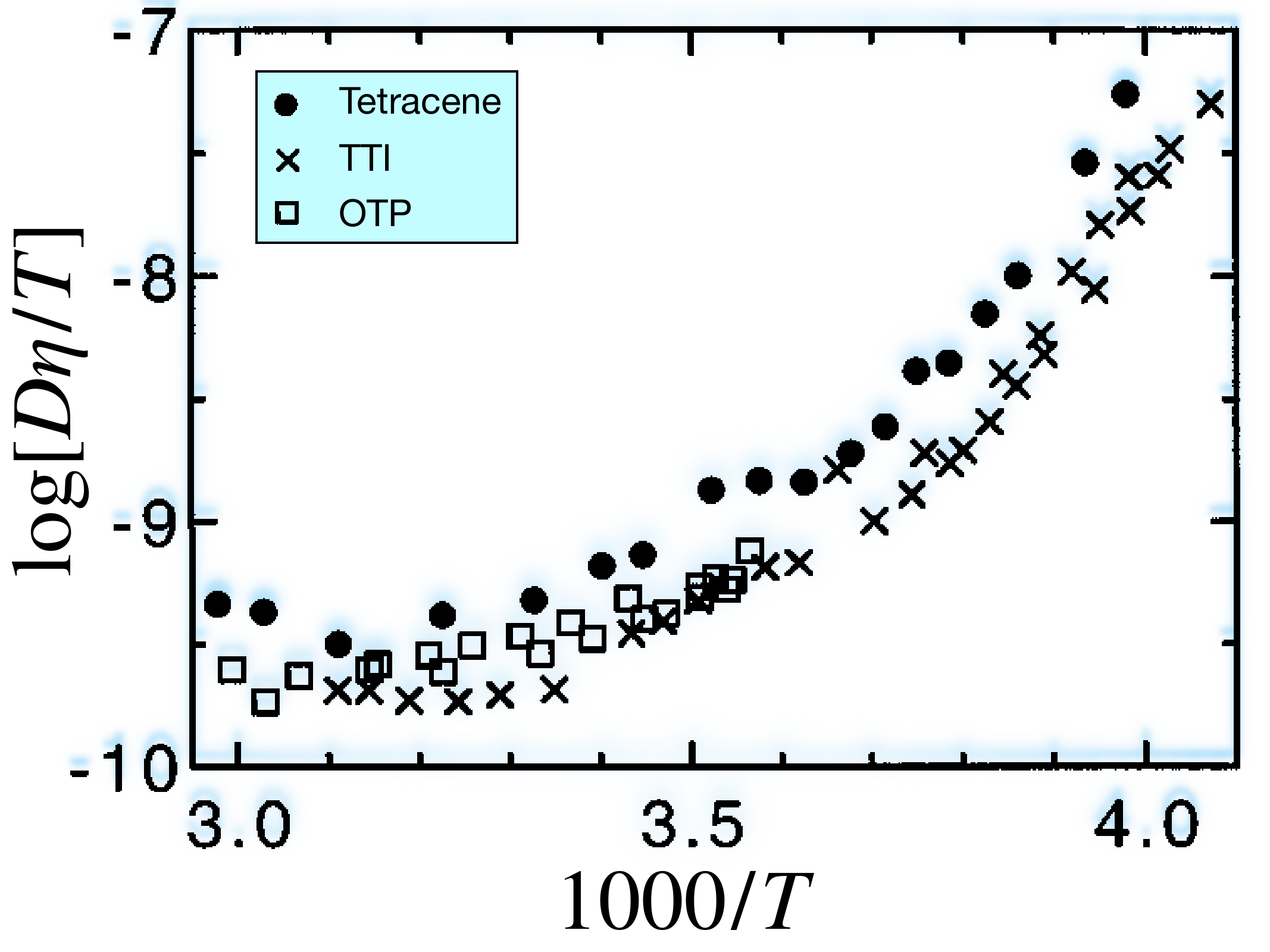}
	\caption{$D\eta/T$ will be constant if the Stokes-Einstein relation is valid. This ratio increases with decreasing $T$ in a glassy system showing the breakdown of the relation. Adapted with permission from Ref. \cite{cicerone1996}.}
	\label{SEbreakdown}
\end{figure}

Another way to determine the properties of particle displacements is to look at the van-Hove function, $G_s(r,t)$. It gives the probability distribution of particle displacement $r$ at time $t$ \cite{Kob1995,Chaudhuri2007,pareek2023}:
 \begin{align}
 G_s(r,t) = \frac{1}{N} \left< \sum^N_{i=1} \delta(r-|\textbf{r}_i(t)-\textbf{r}_i(0)|) \right>.  \label{eq:Gsrt}
 \end{align} 
In the high $T$ liquid phase, when $F_s(k,t)$ is exponential, $G_s(r,t)$ is Gaussian. However, as the system approaches $T_g$, $G_s(r,t)$ deviates from Gaussian. One can also define the van-Hove function along a particular direction $x$, $G_s(x,t)$, as
\begin{equation}
G_s(x,t) = \frac{1}{N} \left< \sum^{N}_{i=1} \delta(x-(x_i(t)-x_i(0))) \right> . \label{eq:Gsxt}
 \end{equation}
 The non-exponential nature of $F_s(k,t)$ is concurrent with the non-Gaussian nature of van-Hove functions (Fig. \ref{vanHove}) \cite{pareek2023,Chaudhuri2007}. One can also characterize the degree of non-Gaussian nature via the non-Gaussian parameter, $\alpha_2(t)$. For example, in spatial dimension 3, one has,
 \begin{equation}
 \alpha_2(t) = \left[ \frac{3\left<r(t)^4\right>}{5\left<r(t)\right>^2} -1 \right]. \label{eq:alpha2}
 \end{equation}

Another intriguing dynamical characteristic of glassy systems is the dynamical heterogeneity (DH). It refers to the coexisting fast and slow-moving regions (Fig. \ref{DH_chi4}a). Moreover, they move in time: a fast-moving region can become slow-moving at later times and vice-versa. A four-point correlation function that characterizes the DH is \cite{Dasgupta1991,franz2000,smarajitPNAS2009,dhbook,IMCT} 
 \begin{equation}
 \chi_4(t) = N \Big[\left<\tilde{Q}(t)^2\right>-\left<\tilde{Q}(t)\right>^2\Big]. \label{eq:chi4t}
 \end{equation}
 where $\tilde{Q}(t)$ is defined in Eq. (\ref{Qoft}). $\chi_4(t)$ increases at short times, attains a peak value, $\chi_4^{\text{p}}$, and then decays again (Fig. \ref{DH_chi4}b). The time when $\chi_4(t)$ has the peak defines another relaxation time, $\tau_\text{peak}$. In general, $\tau_\text{peak}$ is proportional to $\tau$. $\chi_4^{\text{p}}$ is proportional to the average volume of the fast or slow-moving regions. As the system approaches the glass transition point, $\chi_4^{\text{p}}$ increases, signifying DH grows.

Many variables can characterize the transport properties of a system: diffusivity, $D$, viscosity, $\eta$, or relaxation time, $\tau$. $D$ and $\eta$ of a liquid are related via the Stokes-Einstein (SE) relation, $D=k_BT/(c\pi R \eta)$ where c is a constant that depends on dimension, and $R$ is particle diameter \cite{einsteinpaper,Parmar2017}. Using $\tau\propto\eta/T$ \cite{ShiladityaJCP2013}, the SE relation implies $D\eta=$constant or $D\tau=$constant. However, as shown in Fig. \ref{SEbreakdown}, this relation breaks down in the supercooled temperature regimes in the presence of DH \cite{cicerone1996,Parmar2017}. This violation is another characteristic of glassy systems and is found to be directly related to the growing DH.

\begin{figure}
	\includegraphics[width=8.6cm]{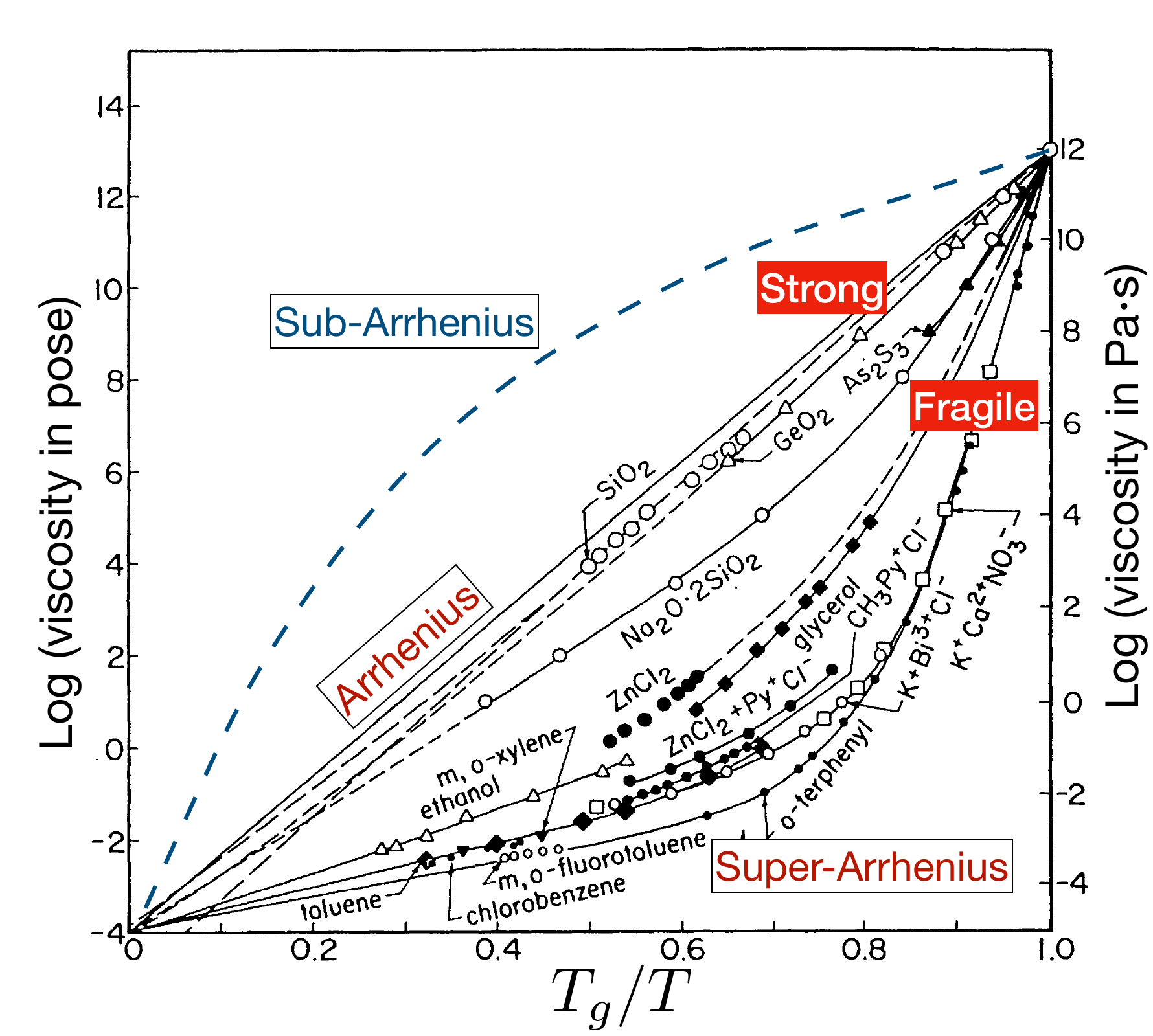}
	\caption{The Angell plot of $\log(\eta)$ as a function of $T_g/T$. Systems whose curves are close to the Arrhenius line are known as `strong' glasses, and away from it, in the lower half, are known as `fragile' glasses. The super- and sub-Arrhenius behaviors are also marked. Adapted with permission from Ref. \cite{angell1995}.}
	\label{angellplot}
	\end{figure}

A steep increase of $\eta$ (or $\tau$) is a defining feature of glasses. However, $\eta$ for different systems will grow at different rates. C. A. Angell showed that the plots of $\log_{10}\eta$ as a function of $T_g/T$ give different curves for various systems (Fig. \ref{angellplot}). The curves meet at $T_g/T=1$ by definition since a specific value of $\eta$ defines $T_g$. This plot is known as the Angell plot \cite{Angell1991,angell1995,Angell1997}. We can categorize various systems as strong or fragile glasses based on the position of the curves in this plot. An Arrhenius behavior, i.e., $\eta\sim \exp[C/T]$, where $C$ is a constant, will follow a diagonal straight line. The systems for which the curves are close to the Arrhenius plot are known as strong glasses, while the systems for which they are away from the Arrhenius plot are known as fragile glasses. Note that the `strong' and `fragile' distinctions are not mechanical. The behavior of the curves for the fragile glasses is known as super-Arrhenius. Likewise, if the curves are on the other side of the Arrhenius line, they are sub-Arrhenius.  For most equilibrium glassy systems at high enough densities, with a few exceptions \cite{berthier2009,berthier2009b,adhikari2022}, the plots are either Arrhenius or super-Arrhenius. \cite{Angell1991,Berthier2011a,Biroli2012,biroli2013}.
The fragility index, $K$, of a system can be defined via 
\begin{equation}
\tau=\tau_0\exp\left(\frac{1}{K(T/T_K-1)}\right),
\end{equation}
where $\tau_0$ is a microscopic time scale. One can fit the above expression with simulation or experimental data and obtain $K$.

\begin{figure}
	\includegraphics[width=8.6cm]{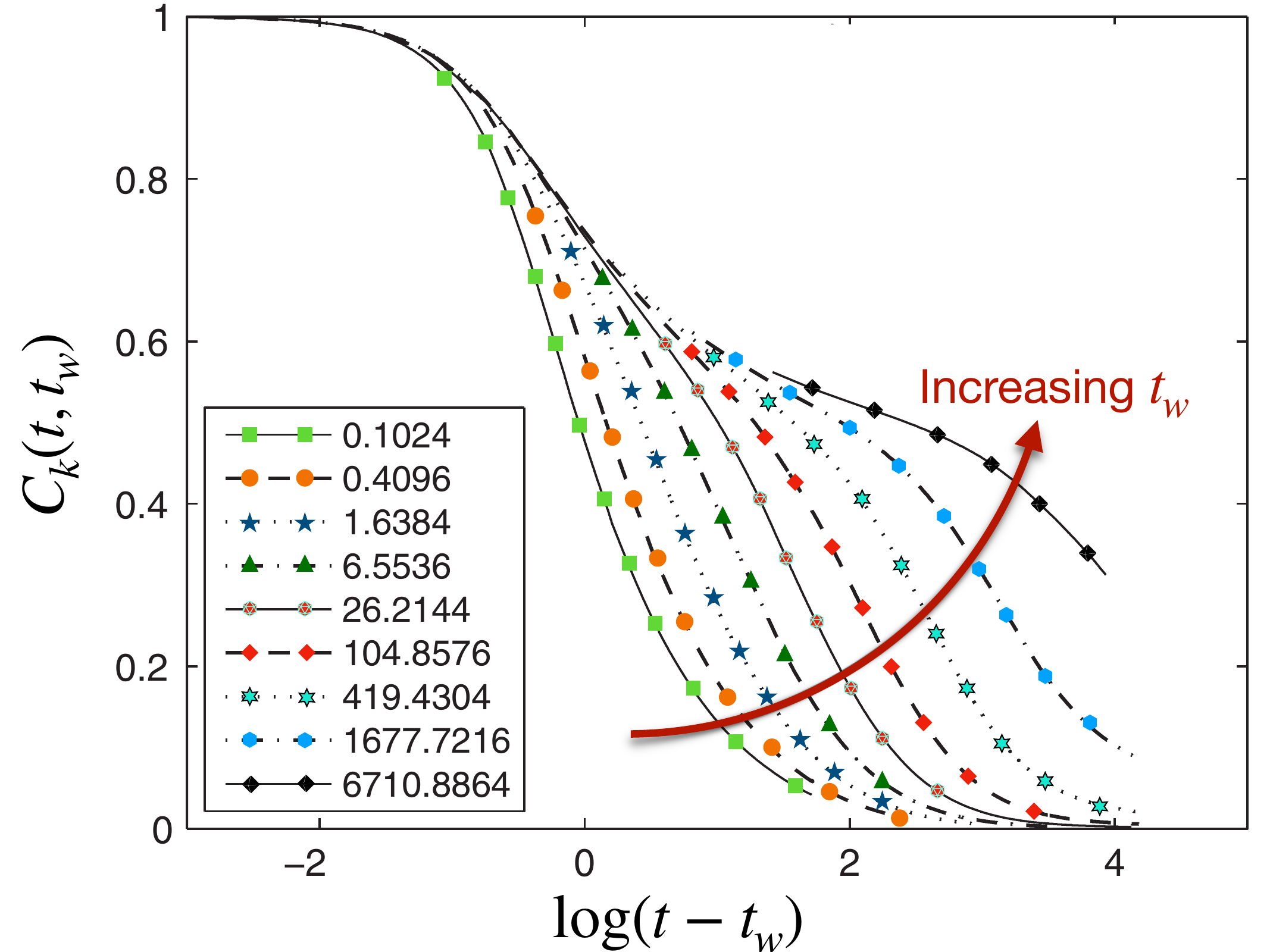}
	\caption{After a sudden quench to a low-$T$, a glassy system evolves towards equilibrium. The waiting time, $t_w$-dependence of the two-point correlation function, $C_k(t,t_w)$, characterizes the non-stationary aging state. (Adapted with permission from \cite{Nandi2012}).}
	\label{aging}
\end{figure}

Finally, we discuss another feature of glassy systems, known as aging. Note that the definition of $T_g$ is motivated by practical considerations than any genuine phase transition. Below a particular time scale, it becomes impractical to keep the system in equilibrium; $T_g$ is the $T$ that corresponds to this time scale. Below $T_g$, the system is out of equilibrium and continues to evolve. After a sudden quench around $T_g$, the system evolves toward the equilibrium state. This non-stationary nature of the state is known as aging: the system properties depend on the age or waiting time, $t_w$ \cite{kob1997,Nandi2012,nandi2016,Debenedetti2001}. For example, the two-point auto-correlation function, $C_k(t,t_w)=\langle\rho_k(t)\rho_{-k}(t_w)\rangle$, depends on both times, $t$ and $t_w$, and not the time-difference alone (Fig. \ref{aging}). Traditionally, the nonequilibrium phase below $T_g$ is called glass, whereas the equilibrium phase above $T_g$ is called super-cooled liquid. However, $T_g$ has no thermodynamic significance.

\section{Experimental results: active glasses}
\label{expts}  

\begin{figure*}
	\includegraphics[width=12.5cm]{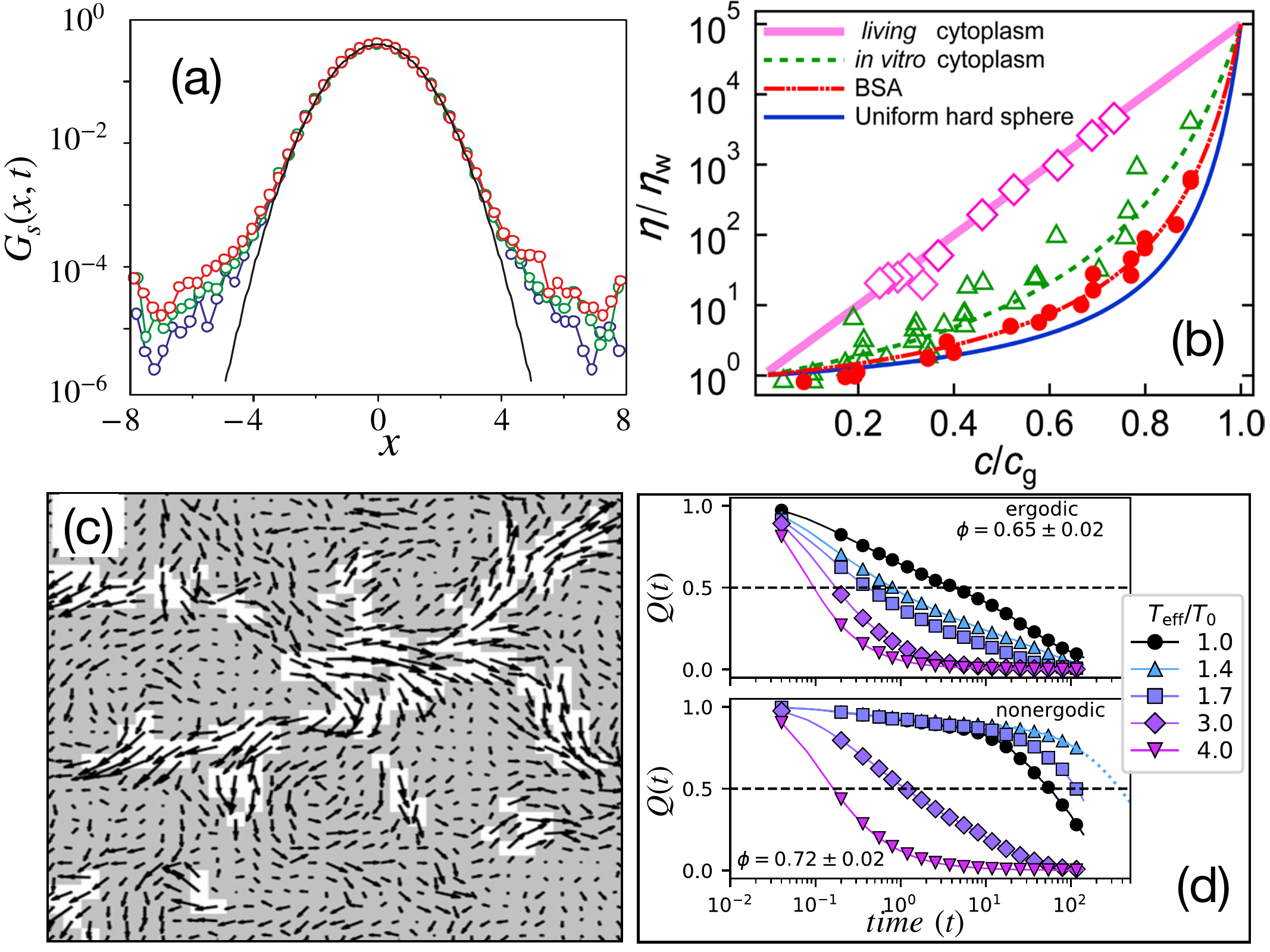}
	\caption{Representation of some experimental results on active glass. (a) The van-Hove function for particle displacements in a particular direction, $G_s(x,t)$, is non-Gaussian for the cytoplasmic fluid. Taken with permission from \cite{bursac2005}. (b) The Angell plot representation of viscosity in the cytoplasm. [Taken with permission from \cite{Nishizawa2017}]. The fragility decreases as activity increases. (c) The velocity snapshot of a cellular monolayer shows dynamical heterogeneity. Taken with permission from \cite{Angelini2011}. (d) The overlap function, $Q(t)$, of a dense active system of Brownian particles shows glassy characteristics. [Adapted with permission from \cite{Klongvessa2019a}].}
\label{experimental}
\end{figure*}

We now turn to active glasses. We start by reviewing some of the experimental results that motivated this field. The systematic effort to reveal glassy dynamics in biological systems began in the early 2000s, around the same time when many of the crucial concepts of equilibrium glasses, such as the dynamical heterogeneity and various static and dynamic length scales, just started to evolve \cite{franz2000,Chaudhuri2007}. Many experimental works have revealed the glassy dynamics in diverse biological systems (see Fig. \ref{experimental} for some representative results). For the constraint of space, we will be brief here and refer the reader to some excellent reviews \cite{Sadati2013,Atia2021} for a more exhaustive list of the experimental works. We aim to highlight the diversity of systems showing glassy dynamics. The list is enormous: cellular cytoplasms, collections of cells and tissues, synthetically designed systems, crowded environments of various organisms - from ants to humans, etc. These experimental results have immensely enriched the field of glassy dynamics.

In the year 2001, the group of Jeffrey J. Fredberg coated ferrimagnetic microbeads with synthetic peptide, bound them to integrin receptors on the surface of human airway smooth muscle cells and showed via rheological measurements that ``{\it the cytoskeleton may be thought of more properly as a glassy material existing close to a glass transition}'' \cite{fabry2001}. In a series of subsequent seminal works, they showed that cell cytoplasm has many glass-like properties. For example, a firmly anchored bead with the cytoskeleton of a living cell shows caging and sub-diffusive MSD at short times \cite{bursac2005,lenorman2007,trepat2007}. The elastic moduli of the cytoskeleton with frequency vary as a power law, but with an exponent smaller than 3/4 that is expected for a reconstituted F-actin system. The exponent $3/4$ signifies entropic behavior in semiflexible polymers. A smaller value of the exponent for the elastic behavior implies the system is closer to a glassy system \cite{deng2006}. Much like an ordinary glassy system \cite{Berthier2011a}, the cytoskeleton fluidizes under oscillatory shear, shows aging behavior, and the distribution of particle displacement is non-Gaussian \cite{bursac2005,Trepat2009}. These results shook the traditional thoughts about cell cytoplasm, where only specific signalling mechanisms were assumed to be consequential. Instead, the cell interior is now visualized as a complex chemical space of soft material where biochemistry, molecular crowding, and various physical forces are inseparable \cite{trepat2007}.

Over the years, many experiments established that the cell cytoplasm of diverse systems shows glassy behavior: for example, the dynamics inside the Hela cells \cite{Aberg2021}, the light-induced active motion of intracellular chloroplasts that becomes glassy under dim light \cite{Schramma2023}, or the pH-induced reversible adaptation between a fluid-like and a solid-like states \cite{Munder2016}. Despite the similarities of their glassy behavior with equilibrium glasses, the cell cytoplasm is inherently different. Various active forces are quintessential in these systems and lead to fundamental differences. For example, unlike in equilibrium glasses, the MSD becomes super-diffusive at long times \cite{bursac2005}. The system properties are highly ATP (Adenosine Tri-Phosphate) dependent. The bacterial cytoplasm also shows characteristics of glassy systems that can vary with the degree of ATP supply \cite{Parry2014}. The cytoplasm of an osmotically compressed cell behaves like a strong glass, and the fragility decreases as the ATP supply increases \cite{Zhou2009}. More recently, Nishizawa {\it et al.} \cite{Nishizawa2017} studied the transport properties in diverse systems, both {\it in vitro} and living cytoplasm, and showed that this behavior of decreasing fragility with increasing metabolic activity (higher level of ATP) is more generic. The conventional control parameters of glassy dynamics are $T$, density, and physical interactions. But given these fascinating discoveries of glassiness in active systems, this picture now has to change to include activity as a crucial control parameter. Activity will drive the system out of equilibrium. When the departure from equilibrium is substantial, one must resort to new tools. But, when this departure is slight, and there is a separation of time-scale, ``{\it the fluctuation-dissipation ideas can still be applied: the slowly changing overall state of the system is considered to be a small perturbation}'' \cite{parisi2005}. In this limit, we can use linear-response like ideas to extend the equilibrium theories of glassy dynamics for active systems \cite{bursac2005,kurchan2005,nandi2018,cugliandolo2011,cugliandolo2019,petrelli2020}.

We have till now discussed the glassy dynamics inside the cell. However, biology is organized at different levels and different length scales. We now discuss some experiments showing glassy dynamics in another length scale, in aggregates of cells. Most experimental systems of cellular collectives are confluent, i.e., the cells fill the entire space. The packing fraction remains constant at all times. By contrast, the packing fraction in particulate systems can vary and be a control parameter. 
One clarification on terminology will be beneficial here. The terms - jamming, and glass - are distinct, with entirely different physics \cite{mari2009,berthier2009,ikeda2012,biroli2013}. The first is a zero-temperature, zero-activity phenomenon, whereas glassiness signifies competition between energy barriers and thermal or active agitation. Most biological systems are active. Strictly speaking, the solid-like slow dynamics should be called glassy dynamics. However, these terms are often used imprecisely in this field \cite{Berthier2019d,Janssen2019Review,Atia2021}, and jamming and glassy dynamics are often interchangeably used while referring to solid-like slow dynamics.

In a pioneering work, Angelini {\it et al.} \cite{Angelini2011} showed that the dynamics in a confluent monolayer of Madin-Darby canine kidney (MDCK) cells is similar to that in a glassy system. The self-diffusivity within the monolayer exhibits non-Arrhenius behavior, and the system shows dynamic heterogeneity, hallmarks of glassy dynamics. Park {\it et al.} \cite{Park2015a,Park2016b} demonstrated that a confluent monolayer of human bronchial epithelial cells (HBEC) also shows sub-diffusive MSD, stretched exponential slow relaxation, and dynamical heterogeneity, much like a glassy system. Garcia {\it et al.}, via the study of the HBEC confluent monolayer, established that the system exhibits a long-range velocity correlation, similar to self-propelled systems in the dilute regime \cite{Garcia2015}. Malinverno {\it et al.} \cite{Malinverno2017} showed that a confluent human mammary epithelial cell monolayer also shows glassy characteristics; they further demonstrated that the system fluidizes when a particular cortical functional protein, RAB5A, is over-expressed \cite{Palamidessi2019}. Different confluent monolayers, such as HBEC and MDCK monolayers, the Drosophila wing disk, etc., also show similar glass-like behavior \cite{Nnetu2012,Sadati2013,Giavazzi2017,Giavazzi2018,Atia2018,Lin2020,Kim2020,Vishwakarma2020}. Schötz {\it et al.} \cite{Schotz2013} revealed that Zebra-fish embryonic explants have glassy properties, such as anomalous diffusion, caging behavior, non-Gaussian particle displacements, etc. Mongera {\it et al.} \cite{Mongera2018} showed the existence of a positive stress gradient from posterior to anterior during the vertebrate body axis elongation in Zebrafish embryos. It correlates with the fluid-like behavior in the posterior zone and the solid-like glassy behavior on the anterior side. A fluid cannot support stress, whereas a solid can. They have shown that active stress fluctuations fluidize the tissue in the posterior zone, and ``{\it cell rearrangements and movements are all consistent with the tissue behaving as a disordered, glassy material}" \cite{Mongera2018}. Thus, one common theme appears via all these experimental results. Irrespective of the detailed cell types, a confluent monolayer can exhibit glass transition and such dynamical behavior is relevant for several biologically significant processes.

We now discuss some examples of glassy dynamics at various other length scales. Bacterial colonies can exist in different phases, such as liquid, glassy, active nematic, etc. As the number density increases, the dynamics within the colony shows a crossover from a swarming state to a slowed-down glassy state \cite{Takatori2020,lama2022}. The aggregation of macroscopic insects such as fire ants also shows remarkable similarities with a glassy system \cite{Gravish2015,Tennenbaum2016,Helbing2000,Helbing2001,Helbing2005,Bain2019}. Research on disease-spreading mechanisms reveals that glassy dynamics of the adaptive immune response to antigens prevent autoimmune diseases \cite{sun2005}. Very recently, several works have also shown that the biomolecular condensates, i.e., the phase separated dense region of intracellular proteins \cite{brangwynne2009,hyman2014}, also show glassy behaviors \cite{jawerth2020,alshareedah2021,lin2022,frank2023}.

Most biological systems are too complex to be amenable to a detailed theoretical treatment. However, we can study different aspects individually by defining simpler model systems with specific characteristics; this has proved immensely powerful in physics and provides deeper insights into complex problems. Synthetically designed model systems mimic various active systems; e.g., symmetric and asymmetric rod-shaped particles on a vibrated disk represent active systems of SPPs \cite{Dauchot2005,Ramaswamy2010,Narayan2007,Marchetti2013}. Arora {\it et al.} \cite{Arora2022} have designed an experimental system consisting of 3$d$-printed prolate ellipsoids on a vertically vibrated plate. Asymmetric friction and a hole along the principal axis of the ellipsoid can precisely control particle activity. The experiments confirm the re-entrance phenomenon of glassy dynamics and the disappearance of glassy dynamics at high enough activity. Synthetic Janus particles with two different surfaces can self-propel in certain fluids \cite{Casagrande1989,Jiang2010,Bechinger2016}. Klongvessa {\it et al.} \cite{Klongvessa2019a,Klongvessa2019b} studied the glassy dynamics in a system of gold Janus particles half-coated with platinum. They showed that the overlap function exhibits two-step relaxation with a plateau at intermediate times, implying caging of the particles. The plateau gets longer and the system becomes non-ergodic with increasing density \cite{Klongvessa2019b}. The relaxation dynamics shows complex stretched exponential relaxation with decreasing activity. 
In a recent work, Arora {\it et al} \cite{arora2024} have introduced a fascinating system to mimic the dynamics of a cellular monolayer. They take a thin paper clip, glue the two ends to form a ring, place the 3$d$-printed active particles inside this paper ring, and place the entire system on a vertically vibrated plate. This system represents a single synthetic cell. Placing several of these ``cells" on the plate, they mimic a cell monolayer. Remarkably, this system reproduces several static and dynamic properties of a cellular monolayer. However, in comparison with biological systems, one has immense control over this synthetic system. Specifically, the results of Ref. \cite{arora2024} demonstrate that the jamming transition and the glassy properties of epithelial systems come from geometric constraints.

The above examples show that glassy dynamics is prevalent in various biological systems at different length scales. These examples have immense practical importance. All these aspects make active glass a fascinating scientific problem. However, due to the inherent complexity of these systems, it is not clear if the mechanisms governing the glassy dynamics in different systems are related. A theoretical approach can help in addressing these questions. However, developing a theory for such systems is certainly non-trivial and challenging; as is often the case, numerical simulations can greatly help in such a scenario.

\section{Simulation study of active glassy systems}
\label{simulations}
As biological systems are immensely complex, simulations have provided crucial insights into their glassy dynamics. We will first discuss the particle-based model systems of SPPs and then the confluent models of epithelial tissues in Sec. \ref{confluentmodels}.

Theoretical implementation of activity in the form of SPPs can be of many different forms; the essential idea is to break the detailed balance such that noise is no longer related to dissipation via the fluctuation-dissipation relation \cite{Ramaswamy2010,Marchetti2013,hansenmcdonald}. Thermal noise is $\delta$-correlated over time. One straightforward way to implement activity is to use a colored noise correlated over time. The correlation time of the active noise is known as the persistence time $\tau_p$. This persistence time is a crucial aspect of active forces. There exist many possible ways to implement activity in the form of self-propulsion. We will only discuss some of the most well-known forms.

\subsection{Different models of self-propulsion}
\label{differentmodels}
The models of self-propulsion, also known as motility, that most simulations have implemented till now are of three broad categories: active Brownian particles (ABP),  run-and-tumble particles (RTP), and active Ornstein-Uhlenbeck process (AOUP). The long-time behaviors of a dense system with respect to varying $\tau_p$ are similar for the first two types of noises and different from AOUP \cite{Nandi2017b,Nandi2018a}.

The Brownian motion refers to the erratic motion of a particle as a result of random kicks by particles of the bath. The equation of motion for the active Brownian particle is
\begin{align}
m\dot{\mathbf{v}}=f_0\hat{\bf n}+\sqrt{2D_T}\bm{\zeta}; \,\,\,
\dot{\phi}=\sqrt{2D_R}{\xi}
\end{align}
where $\mathbf{v}=\dot{\mathbf{r}}$ with $\mathbf{r}$ being the position of the particle, $\hat{\bf n}=(\cos\phi,\sin\phi)$, $f_0$ is the self-propulsion force \cite{howse2007,Bechinger2016}, $\bm{\zeta}$ and ${\xi}$ are the noises of zero mean and unit strength. $D_T$ and $D_R$ give translational and rotational diffusivities. Setting $f_0=0$ provides the equations of motion for passive particles. ABPs undergo random fluctuations and directed active swimming, driving these particles out of equilibrium.

The run-and-tumble particle (RTP) dynamics was originally proposed to describe the dynamics of {\it E. Coli} bacteria \cite{berg1979}. The particles move with a constant speed of $v_0$ and reorient after a persistence time $\tau_p$. The reorientation event is tumble; $\tau_p$ has a Poisson distribution. The long-time properties of ABPs and RTPs are similar. For the active glassy dynamics, the active noise for these types of systems can be written as
\begin{equation} \label{model1}
\langle f(t)\rangle=0; \,\,\, \langle f(t)f(t')\rangle=f_0^2 \exp[-|t-t'|/\tau_p],
\end{equation}
where $f(t)$ is the active noise at time $t$. One can derive this form of the active noise statistics as a coarse-grained form of the microscopic random kicks in the form of shot noise \cite{benisaac2015}.

On the other hand, several works have also included activity as an active Ornstein-Uhlenbeck process (AOUP) \cite{Szamel2016, Flenner2016, Ghoshal2020}. The over-damped equation of motion for the particles comprising such a system is
	\begin{align} \label{model2}
 \dot{\mathbf{r}}_i &= \xi^{-1}_0 [\mathbf{f}_i + \sum\limits^N_{j(\neq i)=1} \mathbf{f}_{ij}]; \nonumber \\ 
	\tau_p \dot{\mathbf{f}_i} &= -\mathbf{f}_i + \bm{\zeta}_i. 	\end{align}
where the $\bm{\zeta}_i$ has zero mean and variance of $2\xi_0 \Teff^\text{sp} \delta_{ij} \delta (t-t')$. $\Teff^\text{sp}$ is the single-particle effective temperature, similar to $f_0^2$, and denotes the strength of the active noise. $\xi_0$ denotes the friction and can be set to unity. The active noise correlation in this case becomes
\begin{equation} \label{model2form}
\langle \mathbf{f}_i(t)\rangle=0; \,\,\, \langle f_{i,\mu}(t)f_{j,\nu}(t')\rangle=\delta_{ij}\delta_{\mu\nu}\frac{\Teff^\text{sp}}{\tau_p} \exp\Big[-\frac{|t-t'|}{\tau_p}\Big],
\end{equation}
where $\mu$ and $\nu$ denote spatial components of the active force.

Although other forms of activity are also possible (see Sec. \ref{othermodels}), these two forms describe most of the active systems. Their forms are motivated by different biological systems. In the first set of models, generally, there are two types of molecules, A and B. Active forces are effective when A’s are attached to the B’s. There is an attachment-detachment dynamics with $\tau_p$ referring to the time scale A remaining attached to B. Naturally, when $\tau_p\to0$, there is no active force; this is easy to verify from Eq. (\ref{model1}). This type of activity is known as model 1 or the SNTC (Shot Noise Temporal Correlation) model \cite{Nandi2017b,Nandi2018a}. By contrast, when activity machinery is internal to the particles, $\tau_p$ refers to the time of rectilinear motion in a particular direction. In this case, activity strength is maximum and the system follows equilibrium Brownian dynamics when $\tau_p\to0$; the activity strength decreases as $\tau_p$ increases. Equation (\ref{model2}) implements this scenario; this type of activity is known as model 2 or AOUP \cite{Nandi2017b,Nandi2018a}. Although the effects of self-propulsion are similar within both models, the trends as a function of $\tau_p$ are opposite \cite{Mandal2016b,Flenner2016,Nandi2017b,Nandi2018a}.

\begin{figure*}
		\includegraphics[width=14cm]{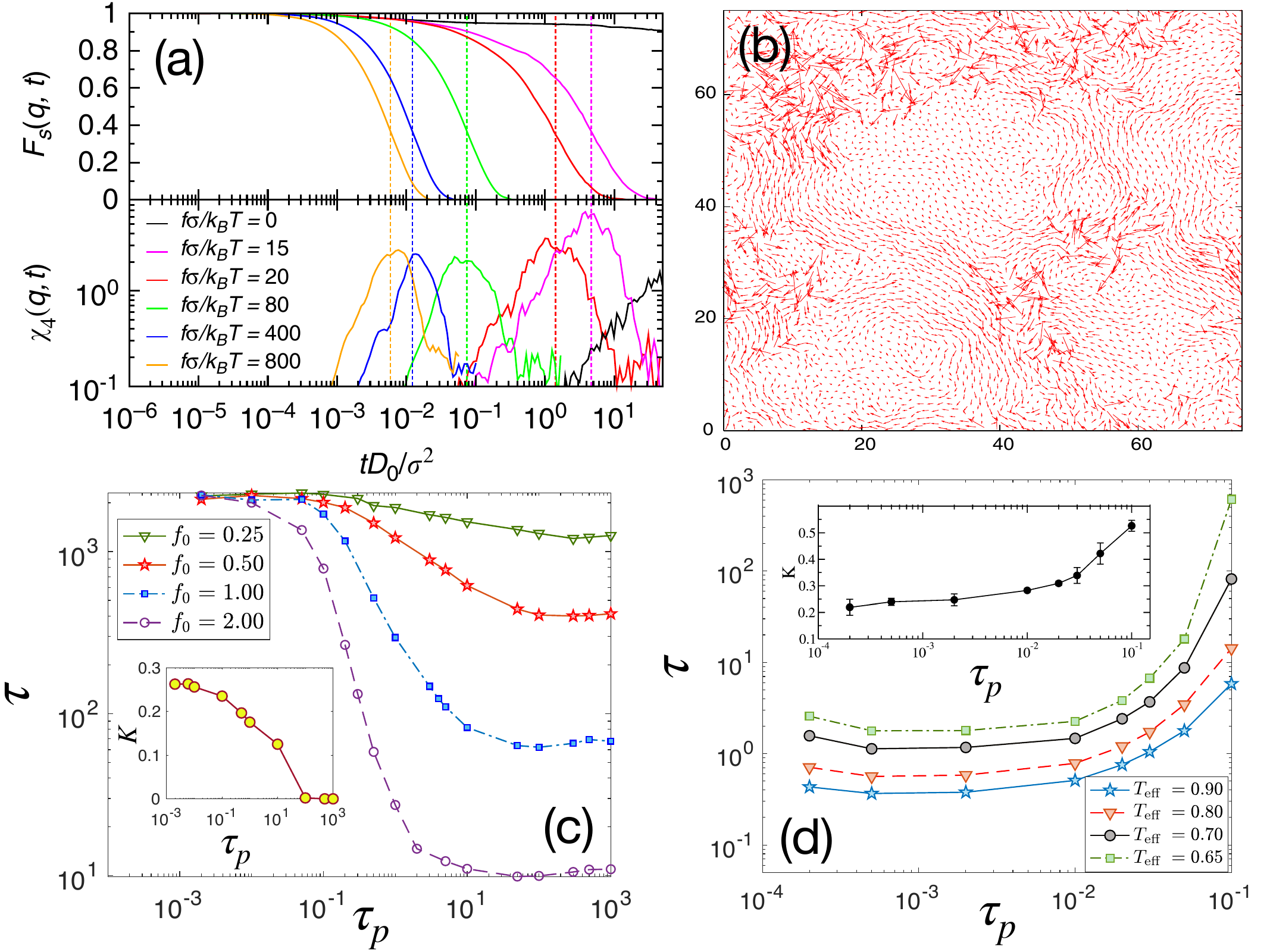}
		\caption{(a) Relaxation becomes slower as activity decreases in a hard-sphere model of ABP. The decay of $F_s(q,t)$ becomes slower, but there is no corresponding growth in the peak of $\chi_4(q,t)$. [Taken with permission from \cite{Ni2013}]. (b) DH increases with decreasing activity in a mode active system. [Taken with permission from \cite{Berthier2014b}]. (c) Relaxation time, $\tau$, and fragility $K$ decreases as $\tau_p$ increases in model 1. Adapted from Ref. \cite{Nandi2018a}. (d) $\tau$ and $K$ increases as $\tau_p$ increases in model 2. Reproduced from Ref. \cite{Flenner2016} with permission from the Royal Society of Chemistry. }
		\label{simulation_fig}
	\end{figure*}

\subsection{Simulations of active glasses of SPPs}
We now summarise some simulation works exploring glassy dynamics in dense active systems of SPPs. Many ``intuitive'' results may prove wrong in active systems. Considering activity as a driving force, it may seem plausible that the glass transition is entirely suppressed due to activity, much like a glass under steady shear \cite{Berthier2019c,Berthier2011a}. In 2011, Henkes, Fily, and Marchetti showed that a system shows glassy behavior even in the presence of activity \cite{henkes2011}. Though the detailed behavior depends on the specific model and the parameters \cite{Berthier2019c}, most simulations show that active driving delays the glass transition. For example, one can reach the universal random close packing fraction (RCP) of $0.64$ in a hard-sphere model by introducing activity in the system. Reference \cite{Ni2013} implemented activity via the ABP model discussed above. As shown in Fig. \ref{simulation_fig}(a), the introduction of activity fluidizes the system, i.e., $F_s(k,t)$ decays faster as activity increases; this allows to equilibrate the system even close to RCP. Surprisingly, the higher-order susceptibility, $\chi_4(k,t)$, shows no increase in peak height, even when relaxation time $\tau$ increases (Fig. \ref{simulation_fig}a). This result contrasts the behavior in equilibrium glasses \cite{Berthier2011a}. Using a slightly different variant of activity belonging to the class of model 2, Berthier showed that a two-dimensional system of self-propelled hard disks undergoes a nonequilibrium glass transition \cite{Berthier2014b}. A comparison with mode-coupling theory (MCT) inspired power-law behavior for $\tau$ as a function of packing fraction $\phi$ provides $\phi_c$, the critical value of $\phi$ where $\tau$ diverges: $\tau\sim (\phi_c-\phi)^{-\gamma}$, where $\gamma$ is an exponent. Similar to Ref. \cite{Ni2013}, $\phi_c$ increases with increasing activity. However, there are crucial differences too. Unlike the results in \cite{Ni2013}, this work suggests a ``re-entrant'' behavior and growing dynamic correlations manifested by the increasing DH (Fig. \ref{simulation_fig}b) \cite{Berthier2014b}. This re-entrance behavior, that is non-monotonic nature of $\tau$ as a function of $\tau_p$, has been revealed by several other works \cite{Berthier2017,debets2021,chaki2020}. However, it is not clear if such non-monotonic behavior is a generic feature of active systems with persistent noise or only appears in specific models of activity.

The effects of activity on the glassy dynamics depend on the details of the active noise. We will discuss two studies by different groups to highlight this. The group of Dasgupta {\it et al.} used a model of self-propelled particles of RTP belonging to the class of model 1 discussed above \cite{Mandal2016b}. Figure \ref{simulation_fig}(c) shows that $\tau$ and fragility $K$ decrease as $\tau_p$ increases. Whereas Berthier, Szamel, and Flenner presented simulation studies of an active system with model 2 type of activity \cite{Flenner2016}. They found ``{\it a very different qualitative picture of the glass transition in systems of self-propelled particles}'' \cite{Flenner2016}: $\tau$ and $K$ increase as $\tau_p$ increases (Fig. \ref{simulation_fig}d). The reason behind the opposite effects of activity is two different models. These two examples highlight the significance of the detailed forms of activity \cite{Nandi2017b,Nandi2018a}.

There are several differences between active glasses and passive glasses. The active systems of anisotropic particles show large swirls or vortices \cite{mandal2017}. Activity governs the scale of these vortices and can become system-spanning as the system approaches glass transition. Activity can either promote or suppress glassy behavior depending on the region of parameter space \cite{Berthier2017}. Interestingly, the active glassy phase correlates well with the two-point static density correlation function \cite{Berthier2017}. This result implies MCT of glassy dynamics should be able to address various features of active glasses. We will show later that this assertion of Berthier {\it et al.} is indeed correct.

Fily, Henkes, and Marchetti have studied the glassy dynamics and phase separation of active systems within the same framework \cite{fily2014}. The phase separation in these systems with repulsive interaction is a novel effect of activity alone. This effect should be there in the dense regime as well. The re-entrance behavior in these systems affirms this effect \cite{Berthier2017,Berthier2014b,Arora2022}. However, this re-entrance in active systems must be distinct from that in equilibrium systems since the effective attractive interaction has a lifetime (changes after $\tau_p$). Exploration of this behavior in detail will provide critical insights into the effects of activity on glassy dynamics. We emphasize that $\tau_p$ is the main activity parameter: the system behavior at small and large $\tau_p$ can be different. This aspect seems relevant even for the aging behavior \cite{Nandi2012} in active glasses \cite{Mandal2020,klongvessa2022}.

\begin{figure*}
		\includegraphics[scale = 0.27]{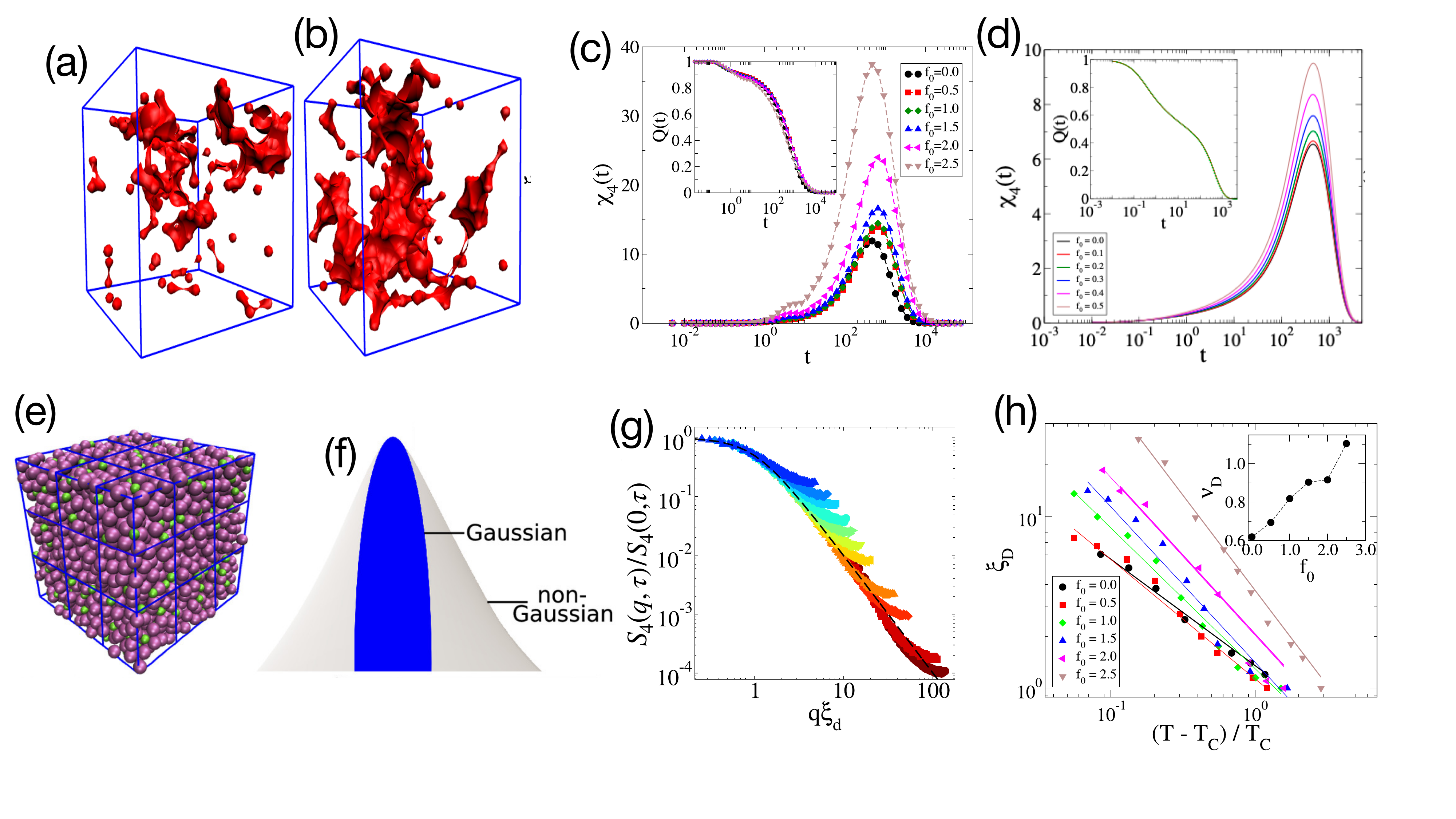}
		\vskip -0.2in
		\caption{(a) Visual representations of growing Dynamic heterogeneity in the system by color coding the set of most mobile particles, often termed as cooperatively rearranging regions (CRR) in the literature for passive systems, (b) shows the same for an active system. Enhanced heterogeneity is evident from the visual representation itself. (c) Four-point dynamic susceptibility, $\chi_4(t)$ as a function of increasing activity parameter $f_0$, keeping the relaxation time of the system the same as depicted in the inset by the decay of the two-point density correlation function $Q(t)$. The dramatic growth of $\chi_4$ peak with increasing activity correctly captures the enhanced DH. (d) $\chi_4(t)$ obtained from active-IMCT also shows similar behavior as in simulation. Inset: As the relaxation time is fixed,  $Q(t)$ overlaps. Within active-IMCT, the parameter $\lambda$ is chosen such that $Q(t)$ overlaps for all values of $f_0$s with that for $f_0 = 0$. (e) Schematic representation of the block analysis method for a finite system size. (f) We can use the non-gaussian nature of the van-Hove function to obtain $\xi_D$. (g) Scaling analysis of the four-point structure factor at the time scale $\tau$, $S_4(q,t=\tau)$ also gives $\xi_D$. (h) Active-IMCT calculation accurately captures the essential physics behind the phenomena of enhanced DH due to activity. Adapted with permission from Ref. \cite{Paul2021a}.}
		\label{activeDH}
	\end{figure*}

In a recent work, Paul {\it et al.} \cite{Paul2021a} have shown that activity has non-trivial effects on the DH. How can we compare the DH of various active systems with varying parameters? Since the relaxation dynamics remains equilibrium-like at a suitably defined $\Teff$, one can choose systems with constant $\tau$ but varying activity and compare their DH to illustrate the role of activity. Figure \ref{activeDH} (a) and (b) show the visual effects of activity on DH as depicted by the cooperatively rearranging region (CRR) (defined as the regions where particles have moved more than the average particle displacement). The cooperative regions grow significantly in size in the presence of activity even if $\tau$ remains the same. Another way to quantify the effect of activity on DH is by measuring the four-point susceptibility, $\chi_4(t)$, as shown in Fig.\ref{activeDH} (c) (simulations) and Fig.\ref{activeDH} (d) (active-IMCT prediction). Notice the dramatic increase of peak height with increasing activity in the simulation results, and the active-IMCT predictions corroborate the same (see the active-IMCT discussion section). The DH length scale, $\xi_D$, plays a central role in various theories of glassy dynamics. In equilibrium systems, $\xi_D$ remains of the order of a few molecular/particle diameters. Thus, the dramatic growth of DH, and consequently large $\xi_D$, in active glasses can be beneficial to test different theoretical predictions more easily.

Since activity drives the system out of equilibrium, measuring $\xi_D$ in these systems is nontrivial. Reference \cite{Paul2021a} measured $\xi_D$ via four different ways to ensure applicability of the methods in nonequilibrium setup: the block analysis \cite{SaurishPRL2017}, via the coarse-graining of the van Hove correlation function \cite{BhanuPRE2018}, the scaling analysis of four-point structure factor \cite{smarajitPNAS2009}, and via the displacement-displacement correlation function \cite{kob1997DH,donati1999} (see Figs. \ref{activeDH} (e)-(g) for a schematic representation). $\xi_D$ obtained from the four methods agree with each other. More recently, similar equilibrium methods of probing $\xi_D$ using elongated probe particles have been extended to these active systems with remarkable agreement amongst them \cite{AnoopPRELett2023}. Having $\xi_D$ as a function of $T$ and activity, one can test various predictions of theories of glass transition. Figure \ref{activeDH}(h) shows $\xi_D$ as a function of scaled temperature $(T-T_{C})/T_C$, where $T_C$ is the MCT critical temperature. The power-law behavior becomes very prominent in active systems, suggesting that active-MCT theories might be good candidates for understanding these systems. 
  
Activity also affects $\xi_D$ at short times \cite{Dey2022}. Dey {\it et al}  argued that activity enhances the phonons leading to higher DH at short times. This result naturally raises the question of the effect of activity in two-dimensional systems where one expects much stronger long wavelength phonon excitation due to the Mermin-Wagner theorem, even in passive systems. Indeed, a recent work \cite{dey2024enhanced} has shown that the Mermin-Wagner theorem gets violated in these systems, and activity makes the long wavelength phonon fluctuations so strong that the Debye-Waller factor in these solids diverges as power-law instead of logarithm in system size. Such an enhancement of long wavelength fluctuations suggests that active particles can probably destabilize solid states in three dimensions. Further studies in this direction will shed more light on the role of fluctuations in the dynamical and thermodynamic properties of active systems. In addition, the monotonic relation between the length and time scale can break down in active glasses. This behavior is also quite different from an equilibrium glass \cite{smarajitPNAS2009}. These results show that DH in active glassy systems is qualitatively different from that in equilibrium systems. A detailed systematic study of this aspect can provide fundamental insights into the DH.

\subsection{More models of activity}
\label{othermodels}

As we emphasized earlier, there can be many forms of activity; we now present some such examples. Physicists got interested in the problem of active matter from the seminal paper by Vicsek {\it et al.} \cite{vicsek1995}, who proposed a minimal model for the ordering transition in two-dimension. Since the model is out of equilibrium, the Mermin-Wagner theorem does not rule out any ordering transition in two-dimension \cite{kardarbook}. The main ingredient of the model is an alignment interaction where each particle tries to align with the average direction of its neighbors with an uncertainty (noise). However, including this interaction in the particle-based models with a hardcore repulsion is computationally costly. Therefore, most studies have implemented the alignment interaction implicitly. Motivated by the experiments of Ref. \cite{Deseigne2010}, Lam {\it et al.} proposed a two-dimensional model of self-propelled hard discs with a coupling between the velocity and the polar axis \cite{lam2015}. Numerical integration at low density reveals the presence of the alignment interaction. A hidden alignment interaction of the Vicsek-like form seems to be a generic feature of many self-propelled active systems \cite{caprini2020}. However, there are differences: the flocking transition in the original Vicsek model is continuous, whereas Lam {\it et al.} find it to be discrete. It is unclear if this difference is significant in the dense regime since the glassy state avoids flocking transition. Similar types of indirect implementation of the alignment interaction have been included in several other works \cite{henkes2011,Giavazzi2018}.

The cellular cytoplasm has many intrinsically disordered proteins (IDP); they can actively change their shape by consuming energy \cite{mikhailov2015}. Shape change can strongly affect the dynamics. Oyama {\it et al.} included this aspect within a simple model where particles can have two different diameters with a stochastic switch rate between the two \cite{Oyama2019}. The simulations show that the system fluidizes with a small volume change accompanied by a change in fragility. Such effects can play crucial roles in the dynamics of bacterial cytoplasm where the force-generating motor proteins are different from those in Eukaryotic cells \cite{Parry2014}. In addition, metabolic activity can also play a critical role in the dynamics, both at the level of proteins, where ATP (adenosine triphosphate) can modulate the interaction strength of IDPs, as well as at the cellular and organism (bacteria) level, where ATP controls the level of self-propulsion. Finally, the dynamics of a system must strongly depend on the interaction. Thus, metabolic activity can be the tuning parameter of glassy dynamics \cite{Nishizawa2017,Parry2014}. We also highlight another form of activity, the attachment-detachment kinetics. One of the proteins that determine the mechanical properties of a cell is the actin filament: it is a long rod-like molecule. It is also a dynamic molecule, where monomers attach in one end and detach from the other end \cite{jacques2015}. This form of activity can also affect the dynamics. One can study another type of active system, initially proposed for the nonequilibrium absorbing phase transitions \cite{daniel2015,daniel2017,galliano2023}. In the $2d$ variant of the model, $N$ disks are randomly placed on a plane. Two discs are active if they overlap; otherwise, they are static. The active discs get a random displacement along the axis connecting the two centers of mass. One can have several variants of this model.

Finally, another form of activity can strongly modulate the dynamics of a system: in the form of division and apoptosis. These two processes are crucial for the growth dynamics of any tissue. Pathogenic conditions appears whenever our body loose control of these two processes. Sinha {\it et al.} \cite{Sinha2020} have analyzed spatially heterogeneous dynamics of cells in an agent-based growing tumor \cite{Malmi-Kakkada2018} spheroid. As we will discuss further in Sec. \ref{glassyconfluent}, including these processes within a simple model is nontrivial due to their immensely complex biological nature. Within the model of Ref. \cite{Sinha2020}, cells grow stochastically in a local pressure-dependent way and divide when they reach a critical size. They implemented apoptosis via a random sudden removal of a cell. The inner cells in the tumor showed slow glass-like sub-diffusive dynamics, whereas cells at the outer layer are super-diffusive. Understanding the essential rules that determine cell division and apoptosis will be crucial for the growth dynamics of tissues.

\subsection{Models of confluent systems}
\label{confluentmodels}
We have till now discussed the glassy dynamics in particulate systems of SPPs. However, tissues and epithelial monolayers are fundamentally different from particulate systems. These cellular systems are confluent, that is cells entirely cover the space. For concreteness of the discussion, we will focus on a monolayer of cells, extension to three dimensions is straightforward. The packing fraction of a monolayer remains unity at all times; hence, it cannot be a control parameter. In addition, the shape of the cells determines most physical behaviors \cite{Sadhukhan2022}. Therefore, including this information within the models is essential for a deeper understanding of these systems. Theoretical models for these systems have been developed and are of great interest for the static and dynamic properties. Although cells are three-dimensional objects, experiments show that the height of a monolayer at a particular stage of development remains nearly the same \cite{Farhadifar2007}. Thus, a two-dimensional description of the monolayer is possible. We will briefly introduce these models and summarize some simulation results for glassy dynamics in such systems.

A theoretical framework for static and dynamic properties of a cell monolayer has two distinct aspects. The first is an energy function, $\mH$, describing the physical properties of a cell, and the second is a confluent model. The cellular cytoplasm behaves like an incompressible fluid \cite{jacques2015}, and the cell height remains nearly the same in a monolayer \cite{Farhadifar2007}. These two properties lead to an area constraint with a target area $A_0$. The simplest way to describe this constraint is an energy cost proportional to $(A_i-A_0)^2$, where $A_i$ is the area of the $i$th cell in the monolayer. The other contribution to the energy function comes from two distinct properties. For most practical purposes, the mechanical properties of a cell come from the cell cortex, a thin layer of cytoplasm just below the cell membrane. The cortex comprises long rod-like molecules known as actin filaments and force-generating myosin molecules. Different cross-linking molecules also contribute to mechanical properties. These molecules try to minimize the cell perimeter. In addition, various junction molecules connect the cortices of the two nearest neighbor cells. Examples include E-cadherin, $\alpha$-Catenin, $\beta$-Catenin, tight junction molecules, etc. They provide adhesive, attractive interactions. Since they are present only at the periphery, their contribution in $\mH$ must be proportional to the perimeter. These two properties lead to an energy cost in $\mH$ proportional to $(P_i - P_0)^2$ where $P_i$ is the perimeter of the $i$th cell and $P_0$ is a constant, known as target area, that parameterizes the intercellular properties. Thus, we can write $\mH$ as
\begin{equation}\label{energyfunction}
\mathcal{H}=\sum_{i=1}^{N} \big[\Lambda_A(A_i-A_0)^2+\Lambda_P(P_i-P_0)^2\big],
\end{equation}
where $N$ is the total number of cells, $\Lambda_A$ and $\Lambda_P$ are elastic moduli related to area and perimeter constraints. $A_0$ and $P_0$ can vary for different cells, but we have kept them uniform for simplicity. We can rescale length by $\sqrt{A_0}$, and write Eq. (\ref{energyfunction}) as
\begin{equation}\label{energyfunction2}
\mathcal{H}=\sum_{i=1}^N\bigg[ \lambda_A(a_i-1)^2+\lambda_P(p_i-p_0)^2\bigg],
\end{equation}
where we have redefined the parameters as $\lambda_A=\Lambda_A$, $a_i=A_i/A_0$, $\lambda_P=\Lambda_P/A_0$, $p_i=P_i/\sqrt{A_0}$, and $p_0=P_0/\sqrt{A_0}$. $A_0$ is the average area when we consider poly-disperse systems.
This energy function can now be evolved at a temperature $T$ with various confluent models. In biological systems, $T$ includes contributions from all possible activities and the equilibrium temperature. Thus, interpretation of $T$ remains unclear, and several definitions of $T$ exist: the ratio of correlation to response function \cite{nandi2018,petrelli2020,Nandi2017b}, from Einstein relation \cite{szamel2014}, etc. Within the theoretical models, $T$ is treated at the same footing as an equilibrium temperature and provides good agreements with experiments \cite{Glazier1993,hirashima2017,Fletcher2014,Park2015a,Park2016b}.

The energy function $\mH$ gives the force on a cell, $\mathbf{F}_i=-\nabla_i\mathcal{H}$. The detailed method to include self-propulsion or motility depends on the particular model, we describe one particular method suitable for the Vertex model (see below for the details). We first assign a polarity vector, $\nhat=(\cos\theta_i,\sin\theta_i)$, where $\theta_i$ is the angle with the $x$-axis. The motile force is $\mathbf{f}_a=f_0\nhat=\xi_0 v_0\nhat$. The friction coefficient $\xi_0$ is generally set to unity. $\theta_i$ performs rotational diffusion \cite{Bi2016}, 
\begin{equation}\label{thetaeq}
\partial_t \theta_i(t) =\sqrt{2D_r} \eta_i(t)
\end{equation}
where $\eta_i$ is a Gaussian white noise, with zero mean and a correlation $\langle \eta_i(t)\eta_j(t^\prime)\rangle = \delta(t-t^\prime)\delta_{ij}$. $D_r$ is the rotational diffusion coefficient, $\tau_p=1/D_r$.

Given the energy function, Eq. (\ref{energyfunction}), and the model of activity, we now need a model for confluent systems for simulations. Many such models exist: some are lattice-based, such as the cellular Potts model (CPM) on square and hexagonal lattices \cite{Graner1992,Glazier1993,Hogeweg2000,hirashima2017}; some are continuum models, such as the Vertex model and the Voronoi model \cite{honda1980,marder1987,Farhadifar2007,Fletcher2014}; then some other models that combine both these aspects, for example, the phase field models \cite{nonomura2012}. All these models use the same energy function, Eq. (\ref{energyfunction}); however, they can differ significantly in their implementation details. These models represent cells as polygons and are inspired by the models of foams \cite{Graner1992,Hogeweg2000,honda1980}. We now provide a brief description of these models.

\begin{figure}
	\includegraphics[width=8.6cm]{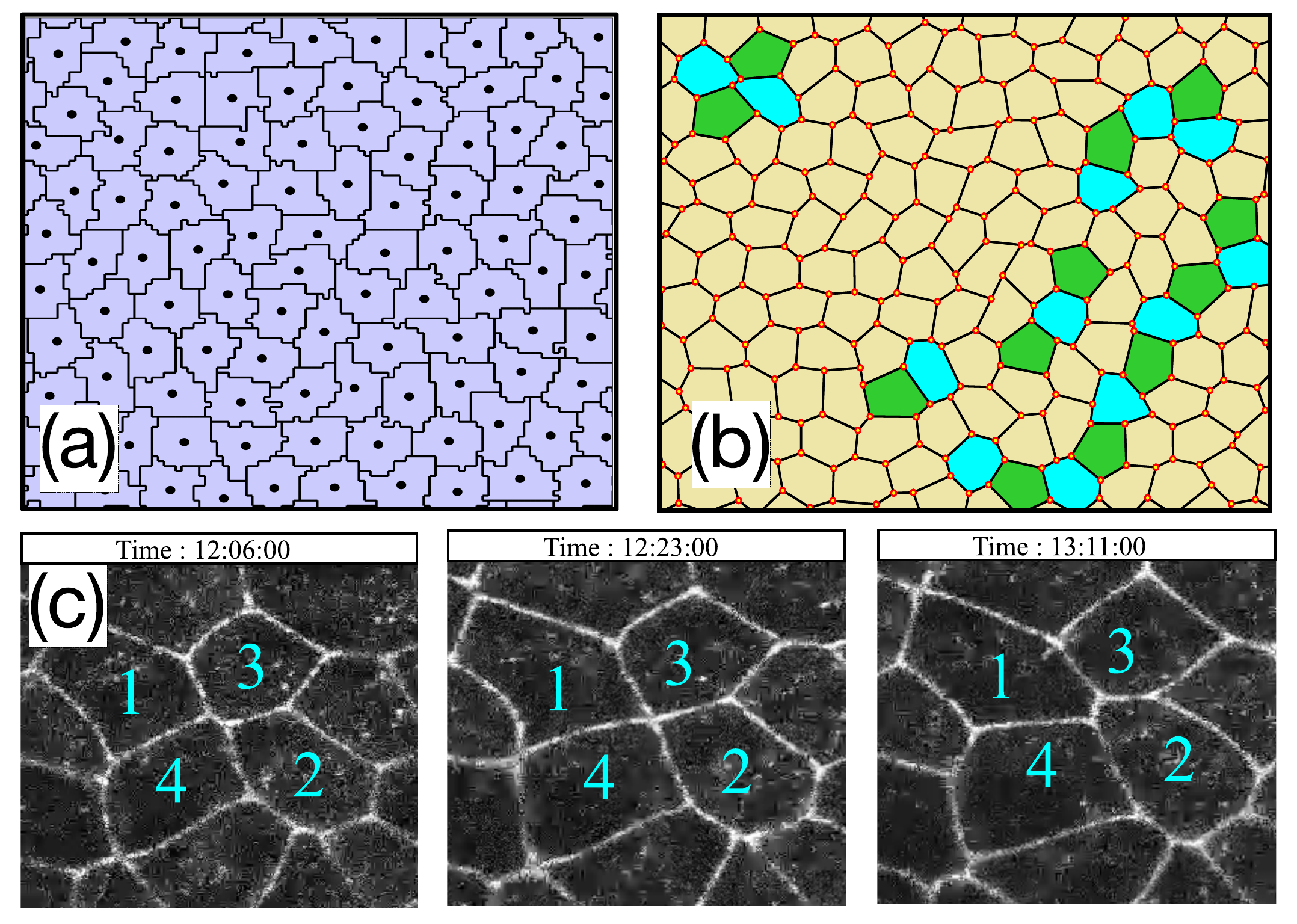}
	\caption{(a) Snapshot of a configuration in cellular Potts model, (b) snapshot of a configuration in Vertex model. (c) Schematic representation of a $T1$ transition. Over time, cell 3 and 4 which were sharing an edge move away, and cell 1 and 2 become the nearest neighbors sharing a newly formed edge under $T1$ transition. [T1 transition snapshots are generated from the Supplementary Movie from Ref. \cite{curran2017}].}
	\label{confluent_schematic}
\end{figure}

{\bf Cellular potts model (CPM):}
The CPM \cite{Graner1992,Glazier1993,Hogeweg2000} is a lattice-based model. Each lattice site has an integer Potts variable $(\sigma)$ from $\sigma \in [1,N]$, where $N$ is the total number of cells. $\sigma = 0$ is usually reserved for medium. The set of lattice sites with the same Potts variable or index represents a cell. The dynamics proceeds via Monte Carlo simulation at temperature $T$. A snapshot of the system from our simulations is shown in Fig. \ref{confluent_schematic}(a). Different cell sites can become disconnected during the dynamics; this is fragmentation. Cells with high activity or $T$ can exhibit such a scenario. However, it is also possible to suppress cell fragmentation via some modified dynamical rules \cite{Durand2016,Sadhukhan2021a}.

{\bf Voronoi and Vertex model:}
In the Voronoi model \cite{Bi2016,Sussman2018b,Paoluzzi2021}, a set of points represent the centers of the cells and are the degrees of freedom. The Voronoi tessellation of these points represents the cells. The cell area and perimeter are those of the tessellated polygons. Dynamics is the evolution of these cell centers either via Monte-Carlo (MC) or molecular dynamics (MD) at a $T$ using the energy function, Eq. (\ref{energyfunction}).
On the other hand, in the Vertex model \cite{Farhadifar2007,Nagai2009,Fletcher2014,Barton2017}, vertices are the degrees of freedom. Figure \ref{confluent_schematic}(b) shows a snapshot of the system, where the vertices are marked. Cell perimeter is defined by connecting the vertices with a straight line (red lines in Fig. \ref{confluent_schematic}b) or a line of constant curvature. Dynamics corresponds to evolving the vertices using the energy function, Eq. (\ref{energyfunction}) either via MC or MD.

In confluent systems, cellular movements proceed via a process known as the $T1$ transition. In the $T1$ transition, cells exchange neighbors. As shown in Fig. \ref{confluent_schematic}, two cells that share an edge move away, and two other cells now share an edge. This process is naturally included within the CPM and the Voronoi models. However, in the Vertex model, it must be included externally: whenever an edge length becomes lower than a threshold value, $\ell_0$, a $T1$ transition is performed.  $\ell_0$ has a crucial effect on the dynamics. This manual implementation of the $T1$ transition can drive the system out of equilibrium. The Vertex model has a rigidity transition, akin to the jamming transition \cite{Bi2015}; however, this transition is absent within the other models \cite{Sussman2018a,Sadhukhan2021a}. Despite this difference, the qualitative dynamic and static behaviors are similar for all three models.


\begin{figure*}
\includegraphics[width=14.6cm]{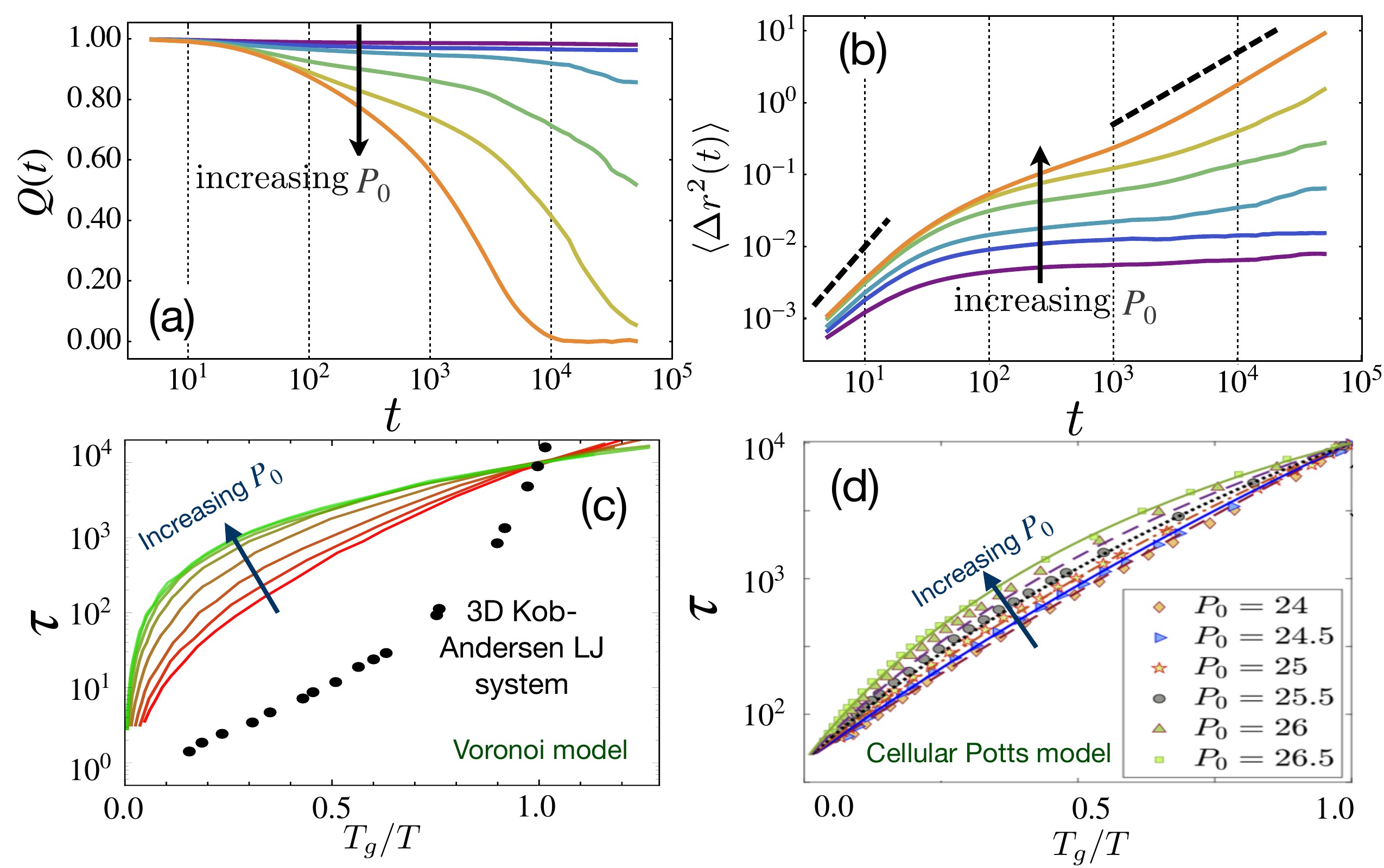}
\caption{(a) The decay of the overlap function becomes faster with increasing $P_0$. (b) MSD also increases as $P_0$ increases. [Taken with permission from Ref. \cite{Bi2016}]. (c) Sub-Arrhenius behavior of $\tau$ in a Voronoi model. [Taken with permission from Ref. \cite{Sussman2018b}]. (d) Sub-Arrhenius behavior of $\tau$ in cellular Potts model. [Taken with permission from Ref. \cite{Sadhukhan2021a}].}
\label{confluent_simulations}
\end{figure*}

\subsection{Glassy dynamics in confluent models}
\label{glassyconfluent}
Glassy dynamics have been investigated via the confluent models both in equilibrium and in the presence of activity. The dynamical behavior within all the confluent models is qualitatively similar. Unlike a system of foam at the confluence, there are many $T1$ transitions in these systems. The distribution of energy barriers for a confluent vertex model is exponential, $\rho(E) \sim e^{-\frac{E}{E_0}}$ \cite{Bi2014}. The dynamics of the system is glass-like, both for $2d$ and $3d$ vertex models \cite{Schotz2013,Bi2014}. In a seminal work, Bi, Yang, Marchetti, and Manning studied a self-propelled Voronoi model and showed that it exhibits a glass transition ``{\it from a solidlike state to a fluidlike state}'' \cite{Bi2016}. The self-intermediate-scattering function shows two-step decay (Fig. \ref{confluent_simulations}a). The MSD grows ballistically at short times, sub-diffusive at intermediate times, and diffusive at long times (Fig. \ref{confluent_simulations}b). From the long-time behavior of MSD, one can define a diffusivity, $D_{\text{eff}}$. Reference \cite{Bi2016} represented the glass transition when $D_{\text{eff}}$ becomes lower than a specific value, $10^{-3}$.  The cell velocity in the presence of activity shows a swirl-like nature, similar to what one finds in asymmetric particles of SPP \cite{mandal2017}. Bi {\it et al.} showed that the glassy dynamics primarily depends on three parameters: self-propulsion speed $v_0$, $\tau_p$, and $p_0$. Similar results were also found in simulations of other models of confluent systems, such as the Voronoi model \cite{Sussman2018a,Paoluzzi2021,Li2021} and the CPM \cite{Chiang2016,Sadhukhan2021a}.

One intriguing property of the confluent systems is a readily-found sub-Arrhenius behavior (Figs. \ref{confluent_simulations} c and d). If we plot viscosity or relaxation time as a function of $T_g/T$, a straight line represents Arrhenius behavior (Fig. \ref{angellplot}). As discussed in Sec. \ref{glassdef}, most equilibrium particulate models exhibit super-Arrhenius behavior. In contrast, confluent systems readily show sub-Arrhenius behavior \cite{Sussman2018a,Li2021,Sadhukhan2021a}. It seems that when the cells are not very stiff, such that $p_0$ is relatively large ($\gtrsim 3$), the system shows sub-Arrhenius behavior, whereas, in the other limit, such that $p_0$ is relatively small, it shows super-Arrhenius behavior \cite{Li2021,Sadhukhan2021a}. However, the origin of this behavior remains unclear.

In a recent study, Paoluzzi  {\it et al.} \cite{Paoluzzi2021} proposed a minimal model for an alignment interaction between the directions of cell elongation and displacement. The strength of this alignment interaction, $J$, governs the glassy behavior and dynamical heterogeneities by forming cooperative regions. $J$ also seems to work as the inverse of an effective temperature; the modified Vogel-Fulcher-Tammann formula in terms of $J$ could capture the structural relaxation time. The glassy dynamics in this model seems qualitatively similar to other confluent systems. These results suggest that the random first-order transition (RFOT) theory might be applicable for the glassy behavior in these systems. RFOT theory has been phenomenologically extended for confluent cell monolayers, and the predictions agree well with both equilibrium and active confluent model simulations \cite{Sadhukhan2021a,Sadhukhan2023}. The simulation results, such as the van-Hove function, $F_s(k,t)$, $\chi_4(t)$, velocity fluctuations, etc, agree well with experiments \cite{Giavazzi2018,Gal2013,Atia2018,Lin2020}.

Close to the glass transition, where the relaxation time is big, the nature of the $T1$ transitions becomes significant. The $T1$ transitions are naturally included within the CPM \cite{Graner1992,Hogeweg2000,Sadhukhan2021a} and the Voronoi models \cite{Sussman2018a,Li2021}. However, it needs to be included externally with some rules within the Vertex model simulations \cite{Fletcher2014,Bi2014}. Although the qualitative behaviors are similar within all the models, there is a crucial difference. A rigidity transition, akin to the jamming transition, has been predicted within the Vertex model \cite{Bi2015}, but no such transition exists within the Voronoi model \cite{Sussman2018a} or the CPM \cite{Sadhukhan2021a}. The influence of this difference on the glassy dynamics remains unclear.

Most studies of the glassy behavior of confluent monolayers do not consider cell divisions and death (or apoptosis). But they are crucial for many biological systems and significantly affect the dynamics. Cell division and apoptosis always fluidize a confluent tissue \cite{Ranft2010,Fernandez2017a,silke2017}, but these studies were within particulate models. Czajkowski {\it et al.}  \cite{Czajkowski2019}  addressed the question using the active Vertex model (AVM). Both cell division and apoptosis are complex biological processes involving many concerted events of intricate natures. Thus, devising straightforward rules to include them within a theory remains challenging. Reference \cite{Czajkowski2019} chose simple rules for these processes, dividing a randomly chosen cell with an arbitrary division plane at a rate similar to apoptosis. For apoptosis, $a_0$ and $p_0$ are set to zero for a cell. Similar rules have also been used elsewhere \cite{Farhadifar2007}, including by some of us \cite{Sadhukhan2022}. A comparable rate for the two processes ensures the conservation of the total number of cells. Contrasting earlier studies \cite{Ranft2010,Fernandez2017b}, Ref. \cite{Czajkowski2019} showed that glassy dynamics exist in a confluent system when the division and death rates are low. Understandably, these processes will strongly affect the other cells in a confluent system. Therefore, a thorough understanding of the rules of these two crucial processes and how they affect different properties of a confluent system is imperative for a deeper understanding of static and dynamic properties of such systems.

\section{Theoretical approaches}
\label{theory}
As discussed in the introduction, the fundamental mechanism of equilibrium glassy dynamics remains unknown. Therefore, applying theories of equilibrium glasses to scenarios in the presence of activity is challenging. However, given the importance of the problem and the presence of a vast amount of experimental data, even the approximate theories are of vital importance for insights. The primary motivation here is to understand the role of activity in systems significantly different from those that physicists usually deal with. Concurrently, these systems provide an opportunity to extend the scope and extent of the original problem. Activity has many forms: the constituent objects can change shape, divide, die, change interaction or valency, self-propel, etc. Significant theoretical development has occurred in the last few years for systems of SPPs and confluent systems. Activity drives the system out of equilibrium. Although the formal proofs fail and new properties emerge, this alone is not the primary difficulty. When ``{\it the departure from equilibrium is substantial, we must resort to different tools... But the situation is different for systems that are only slightly out of equilibrium... In such systems we can expect a separation, by many orders of magnitude, between the microscopic time scale and the macroscopic time scale... The system can then be considered to be essentially thermalized inside a metastable state, and so fluctuation-dissipation ideas can still be applied}'' \cite{parisi2005}. Even though hard to prove analytically, many nonequilibrium behaviors of disordered systems can be explained within a fluctuation-dissipation relation (FDR) framework that is a generalization of the Boltzmann statistics \cite{parisi2005,kurchan2005}. Thus, the main question is how far active systems are from equilibrium. It has been addressed for systems of SPPs, and it turns out not too far when $\tau_p$ is not too large \cite{Fodor2016}. The system still obeys a generalized FDR at a $\Teff$. On the other hand, we can apply mode-coupling theory (MCT) to a non-equilibrium system. We will first discuss this theory and then the generalization of random first-order transition (RFOT) theory for active systems.

\subsection{Mode-coupling theory of active glasses}
Mode-coupling theory is one of the most popular theories of glassy dynamics. It was developed in the early '80s by G{\"{o}}tze and others \cite{gotzebook,Das2004,reichman2005}. It provides an equation of motion for the intermediate scattering function, $F(k,t)$. For a bulk system, the equation of motion is
\begin{align}\label{MCTeq}
\f{\d^2 F(k,t)}{\d t^2}+&\f{k^2k_BT}{S_k}F(k,t)+\int_0^t m(k,t-t')\f{\d F(k,t)}{\d t}=0, \nonumber\\
m(k,t)=&\f{\rho k_BT}{16\pi^3}\int d^3 q|V_{q,k-q}|^2F(q,t)F(k-q,t),
\end{align}
where $k$ and $q$ are wavevectors, $S_k$, the static structure factor, $k_B$, the Boltzmann constant, and $\rho$, the density; note that we have set the particle mass to unity. $m(k,t)$ is known as the memory kernel, and $V_{q,k-q}$ is the vertex function: $V_{q,k-q}=[\hat{q}\cdot {\bf{k}}c_k+\hat{q}\cdot({\bf{k}}-{\bf{q}})c_{k-q}]$ with $\hat{q}$ being the unit vector and $c_{q}$ being the direct correlation function. Equation (\ref{MCTeq}) is an integro-differential equation that we can solve numerically. We can calculate the other variables, such as $\tau$ and $\eta$, via $F(k,t)$. The theory for particulate systems makes several predictions that agree with simulations and experiments \cite{Kob1995,Das2004,gotzebook}. $F(k,t)$ at high $T$ (or low density) decays exponentially. As $T$ decreases, $F(k,t)$ develops a two-step relaxation: it first relaxes towards a plateau and then towards zero at long times, much like in simulations and experiments. As $T$ decreases, the plateau length increases. Eventually, below a particular temperature known as $T_{\text{MCT}}$, $F(k,t)$ remains stuck at the plateau and does not decay to zero: this is a genuine phase transition, known as the non-ergodicity transition or the MCT transition. However, no such transition exists in simulations or experiments, and all the predictions of MCT break down at this point. $T_{\text{MCT}}$ is higher than $T_g$, so the breakdown of the theory happens at a relatively high temperature. The reason behind this failure of MCT remains unclear.

Despite this failure, MCT has several fascinating features for which the theory remains immensely popular \cite{gotzebook,Das2004}. Most simulations and colloidal experiments operate in a parameter space where MCT remains valid. In the regime of validity, the MCT predictions agree well with simulations and experiments. Like a critical theory, MCT predicts power-law divergences for the time and length scales. The exponents are universal and independent of system details. This particular feature of universality makes MCT a natural choice to apply for experimental data of novel systems. MCT assumes that the statics is already known. From the static properties as input, the theory provides the dynamics. One can also write down MCT for nonequilibrium systems \cite{Nandi2012}. We emphasize this specific feature of the theory: the static properties alone provide the dynamics. In active glass simulations, it has been shown that the dynamics is intimately related to the static properties \cite{Berthier2017}. Thus, we expect MCT to work well for these systems.

Concurrent with this expectation, many different variants of MCT exist for active systems of self-propelled particles \cite{Kranz2010,Gholami2011,Berthier2013,Szamel2015b,Szamel2016,Flenner2016,liluashvili2017,feng2017,Nandi2017b,Janssen2018,szamel2019}. Kranz {\it et al.} obtained the MCT for the dynamics of a driven dissipative hard sphere system \cite{Kranz2010}. This model represents synthetic active systems. The theory predicted that glass transition persists even to a high degree of driving. Interestingly, the theory also predicted a weak dependence of MCT exponents on the driving amplitude. The qualitative predictions seem to agree well with simulations of vibrated disks \cite{Gholami2011}. In 2013, Berthier and Kurchan derived an MCT for active spin-glass systems of $p$-spin spherical spins \cite{Berthier2013}. The structure of the theory for this system is similar to that of structural glasses. The authors first write down the theory for a general nonequilibrium state and then demonstrate the conditions when the system resembles an equilibrium system. They showed that ``{\it the main features of this equilibrium glass transition robustly survive the introduction of a finite amount of non-thermal fluctuations driving the system far from thermal equilibrium}" \cite{Berthier2013}. 
Szamel {\it et al}. \cite{Szamel2015b} obtained an analytical theory for the steady state of an active system. The form of the theory resembles that of equilibrium MCT. However, there are crucial differences: the direct correlation function in the memory kernel is replaced by another that combines the velocity correlator, $\omega_{\parallel}$. This difference is a significant departure from the usual MCT as the theory now requires the input of $S_k$ and $\omega_{\parallel}$. Crucially, the spatial correlation of velocities affects the memory kernel \cite{Szamel2016}. Feng and Hou presented an MCT for similar systems where activity enters as AOUP \cite{feng2017}. Liluashvili, {\'{O}}nody, and Voigtmann presented a mode-coupling theory for active systems based on the integration through transients (ITT) approach \cite{liluashvili2017}. ITT has been immensely successful for sheared glassy systems \cite{fuchs2002,fuchs2003}, then it is logical to apply this formalism to obtain MCT for active systems. The qualitative predictions of the theory agree well with simulations \cite{liluashvili2017}. Reference \cite{debetsJCP2022} used the projection operator formalism to obtain the MCT for active systems that has very similar structure as that in Ref. \cite{liluashvili2017} Note that the memory kernels of Refs. \cite{liluashvili2017}, \cite{debetsJCP2022}, and \cite{feng2017} do not include any velocity correlators and the structures are similar to the equilibrium MCT; this contrasts the theory of Refs. \cite{Szamel2015b,Szamel2016}. Unlike in equilibrium, different approaches to deriving the active MCT do not lead to the same final theory. Perhaps, this is not surprising as the system is complex, and the detailed theoretical approach is critical.

\begin{figure*}
\includegraphics[width=14cm]{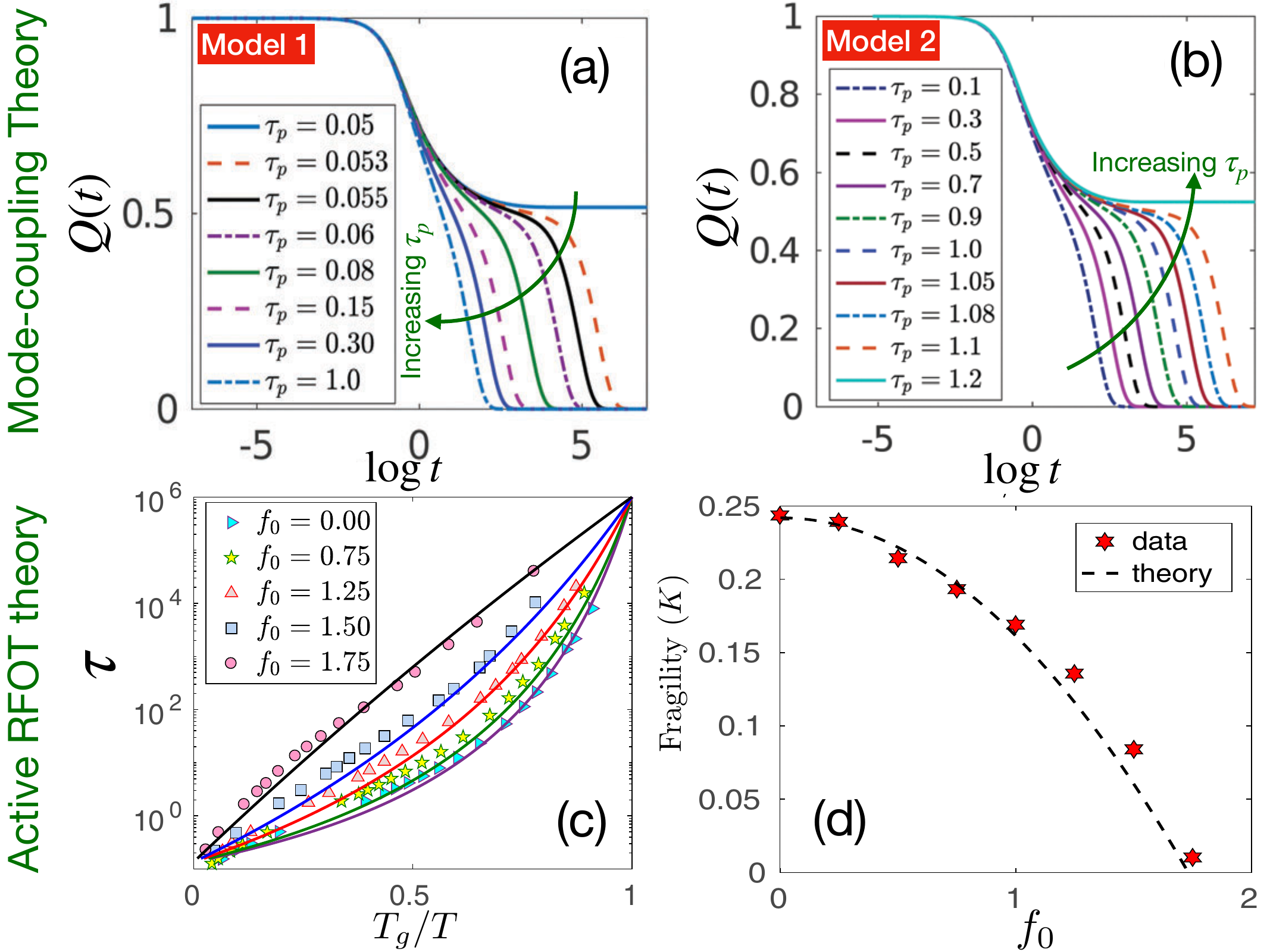}
\caption{(a) Non-equilibrium mode-coupling theory correctly predicts that the system fluidizes when $\tau_p$ increases for model 1 type of activity. (b) The opposite happens for the model 2 types of activity. [Reproduced from Ref. \cite{Nandi2017b} with permission from the Royal Society of Chemistry.]. (c) Comparison of active RFOT theory with simulation data for model 1. (d) Active RFOT also predicts the trend of fragility ($K$) as a function of self-propulsion force, $f_0$. [Taken with permission from Ref. \cite{Nandi2018a}].}
\label{theorymctrfot}
\end{figure*}

The steady state of an active system is out of equilibrium. As Berthier and Kurchan demonstrated for active spin-glass systems \cite{Berthier2013}, a general theory must be in terms of both the response and the correlation functions. As a limiting case, one can write the MCT for the correlation function alone. In equilibrium, FDT ensures this limit is unique. However, no such relation exists for active systems, and the approximation is nontrivial. Possibly, this explains why so many different variants of MCT exist, and their detailed analysis may bring further insights into various MCT approximations themselves. In Ref. \cite{Nandi2017b}, some of us derived an MCT for the steady state of an active glassy system of SPPs via a different route. We first wrote down the most generic theory for a nonequilibrium system, even under aging. We then take the limit of infinite waiting time. In the presence of activity, the system will reach a stationary state. We thus obtain the nonequilibrium MCT for the steady state of active systems. Since there is no FDT-type relation within the derivation, we expect the theory to be valid for the general nonequilibrium steady-state. However, the price one must pay is that it becomes in terms of both the correlation and response function \cite{Nandi2017b}. The schematic version of the theory, written for a particular wavevector, is
\begin{align} \label{activemct1}
\f{\p C(t)}{\p t}&=\Pi(t)-(T-p)C(t)-\int_0^t m(t-s)\f{\p C(s)}{\p s}\d s, \\
\f{\p F(t)}{\p t}&=-1-(T-p)F(t)-\int_0^t m(t-s)\f{\p F(s)}{\p s}\d s, \label{activemct2}
\end{align}
where $C(t)$ and $F(t)$ are the correlation and the integrated response functions. $m(t-s)=2\l\f{C^2(t-s)}{\Teff(t-s)}$, $p=\int_0^\infty \DD(s)\f{\p F(s)}{\p s}\d s$,  $\Pi(t)=-\int_t^\infty \DD(s)\f{\p F(s-t)}{\p s}\d s$, and $\l$ is the control parameter.
$\Delta(t)$ is the variance of active noise, and $\Teff(\tau)$ is defined via a generalized fluctuation-dissipation relation (FDR) for non-equilibrium systems \cite{cugliandolo2011,cugliandolo2019,nandi2018} as 
\begin{equation}\label{Teff}
\f{\p C(t)}{\p t}=\Teff(t)\f{\p F(t)}{\p t}.
\end{equation}
Using simple arguments, Ref. \cite{Nandi2017b} derived an analytical expression for $\Teff$ that agrees well with simulations \cite{Nandi2017b,nandi2018}. Furthermore, they obtained the scaling relations for the relaxation dynamics for both types of active forces discussed in Sec. \ref{differentmodels}; the trend of fluidization as a function of $\tau_p$ are opposite within the two models (Fig. \ref{theorymctrfot} a and b). Consistent with most works, it seems that the relaxation dynamics remains equilibrium-like at a $\Teff$. 
However, as discussed above (Fig. \ref{activeDH}), a recent work have shown that activity has nontrivial effects on the dynamical heterogeneity (DH) \cite{Paul2021a}. Thus, although the relaxation dynamics is equilibrium-like, DH in a glass-forming liquid has qualitatively different behavior. For example, the peak value of $\chi_4(t)$ can vary for the same system with varying activity and $T$ but the same relaxation time. Thus, the DH length scale may have a complex character in active glassy systems. Using two different models, Ref. \cite{Paul2021a} showed that the conclusions are independent of system details. Consistent with existing results \cite{Kranz2010}, this current study also found a weak activity dependence of the MCT exponents \cite{Paul2021a} [see inset of Fig. \ref{activeDH}(h)]. 
Although MCT, till now, has been extended for particulate systems alone, very recently, some of us have applied MCT to the dynamics of confluent systems \cite{pandey2023}. It seems that the unusual glassy dynamics of confluent systems might be an ideal candidate for the MCT-like mechanism of glassiness.

\subsection{Random first-order transition (RFOT) theory}
The random first-order transition (RFOT) theory \cite{Kirkpatrick1987} is another popular theory of glassy dynamics. Through a set of simple-looking arguments, RFOT theory makes many impressive predictions that agree well with simulations and experiments \cite{Kirkpatrick1987,Kirkpatrick1989,Lubchenko2007,Biroli2012,Kirkpatrick2015}. According to RFOT theory, a supercooled liquid comprise mosaics of local states. The free energy of a typical mosaic of size $R$ has two contributions: an energy cost from the interface with other mosaics and an energy gain from the bulk. Therefore, the change in free energy is
\begin{equation*}
	\Delta F = -\Omega_d f R^d + S_d \gamma R^\theta,
\end{equation*}
where, $\Omega_d$ and $S_d$ are volume and surface of a unit hypersphere in dimension $d$, $f$ is the free energy per unit volume, and $\theta$  is an exponent. In general, $\theta \leq d-1$. Minimizing the free energy gives the typical mosaic length scale $\xi$. Now, relaxations within the RFOT theory are entropic. Therefore, we use $f=TS_c$, where $S_c$ is the configurational entropy. The surface energy, $\gamma$ is proportional to $T$, i.e., $\gamma=\Xi T$.

The basis of the RFOT theory is a crucial assumption that $S_c$ goes to zero at a finite temperature $T_K$. In 1948, Walter Kauzmann plotted the ``{\it differences in entropy between the supercooled liquid and crystalline phases}" \cite{Kauzmann1948}, equivalent to $S_c$, for different materials. When extrapolated, the curves for various systems go to zero at a finite temperature \cite{Kauzmann1948}. This surprising result led to the speculation of a finite-temperature phase transition in glassy systems. The phase transition is characterized by a vanishing $S_c$ at $T_K$. Then we can expand $S_c$ around $T_K$: $S_c = \Delta C_p \frac{(T - T_K)}{T_K}$, and obtain 
\begin{equation}\label{eq:lengthscale2}
\xi = \Big [ \frac{\Gamma}{T - T_K} \Big ]^{\frac{1}{d-\theta}},
\end{equation}
where $\Gamma = \frac{S_d T_K \theta \Xi }{\Omega_d d \Delta C_p}$. Relaxation dynamics within RFOT theory involves the relaxation of the mosaics. The barrier height for a region of length $\xi$ is proportional to $\xi^\psi$, where $\psi$ is another exponent. Considering the proportionality constant given by the thermal energy scale, $k_BT$, and assuming a barrier crossing scenario, we obtain $\tau = \tau_0 \exp(\frac{\Delta_0 \xi^\psi}{T})$, where $\Delta_0=k_BT$. The values of the exponents continue to be debated; one possible choice is $\psi = \theta = d/2$ \cite{Kirkpatrick1989,Biroli2012}. Substituting Eq. \ref{eq:lengthscale2} in the expression of $\tau$ and simplifying, gives
\begin{equation}\label{eq:timescale2}
\ln\Big(\frac{\tau}{\tau_0}\Big)= \frac{S_d T\theta}{ \Omega_d d T}\f{\Xi}{S_c} = \frac{\Gamma}{T - T_K},
\end{equation}
where we have set $k_B$ to unity. The predictions of the theory agree well with simulations and experiments. 

The RFOT theory of glass is deceptively simple. Some of its assumptions have deep and profound roots and remain unclear to date \cite{Biroli2012}. Yet, the final expressions of the theory are surprisingly simple and easy to compare with experiments \cite{Lubchenko2007,Kirkpatrick2015}. This feature makes an extension of the RFOT theory for active systems, even if approximate, quite attractive to analyze the data for even more complex systems. Of course, the approximations are nontrivial, but such extensions have provided crucial insights and allowed a platform to think about an exciting problem for fascinating discoveries. We now discuss such extensions of the theory for active glasses.

{\bf Active RFOT theory for systems of SPPs:}
The RFOT theory of equilibrium glasses has been extended for systems of SPPs \cite{Nandi2018a}. Active systems can be considered at an effective equilibrium at a suitable $\Teff$ when $\tau_p$ is not too large \cite{Fodor2016,Szamel2014a,petrelli2020}. Nonequilibrium MCT shows that $\Teff$ is the same as the equilibrium $T$ at short times and goes to a higher value, determined by activity, at long times. The transition from $T$ to the higher value happens at $\tau_p$. Moreover, $\Teff$ explains the relaxation dynamics \cite{Nandi2017b}. These results suggest an effective equilibrium extension of RFOT theory for active systems is possible, at least when $\tau_p$ is not too large. Reference \cite{Nandi2018a} extended the RFOT theory treating activity as a small perturbation and using linear-response-like ideas.

Within RFOT theory, the glassy properties are manifestations of a genuine phase transition at $T_K$, where the configurational entropy vanishes. Notice the behavior of $\tau$, Eq. (\ref{eq:timescale2}): the surface energy appears in the numerator and $S_c$ in the denominator. Since the latter vanishes and the former does not, the critical properties will be dominated by the behavior of $S_c$ close to $T_K$. If the surface energy has no anomalous behavior, one can assume $\Xi$ remains unaffected by activity. However, in active systems, surface energy can have nontrivial behavior. For example, Ref. \cite{bialke2015} numerically studied a model of suspended self-propelled particles and reported a negative interfacial tension. The interfacial tension is not the same as the surface energy of RFOT theory, but they are related \cite{Lubchenko2007}. If the interface tension is negative, one must be careful about the surface term for active systems. However, in a more recent theoretical work, Hermann {\it et al.} challenged the results of Ref. \cite{bialke2015} and analytically showed that the interfacial tension in active systems is actually non-negative \cite{hermann2019}. Negative surface tension will make interfaces unstable; the non-negative value is consistent with the observation of stable interfaces in phase-separated active systems \cite{hermann2019}.

Reference \cite{Nandi2018a} assumed that the activity correction to the surface energy term is negligible and focused on the configurational entropy. When the activity is small, one can expand $S_c$ around its passive value using a Taylor series expansion. The effect of activity is parameterized as a potential $\delta \Phi$ on top of the passive system potential $\Phi$, thus $S_c(\Phi+\delta\Phi)\simeq S_c(\Phi)+\frac{\p S_c}{\p \Phi}|_{\delta\Phi=0}\delta \Phi+\ldots =S_c(\Phi)+\kappa_a \delta\Phi+\ldots$. Then, the expression of relaxation time from the length scale after minimizing the free energy becomes
\begin{equation}\label{eq:timescale3}
\ln\Big(\frac{\tau}{\tau_0}\Big)=  \frac{E}{T - T_K + \frac{T_K \kappa_a \delta \Phi}{\Delta C_p}} ,
\end{equation}
where $E$ is a constant \cite{Nandi2018a}. Therefore, activity shifts $T_K$ to a lower value (compared to the passive case) where  $\tau$ diverges. $\delta \Phi$ is the effective potential due to activity. Within some simplifying mean-field assumptions, we can calculate this contribution for both types of activity discussed in Sec. \ref{differentmodels}. Reference \cite{Nandi2018a} showed that one gets $\delta \Phi = { f_0^2 \tau_p}/(\gamma + k\tau_p)$ for model 1. Substituting it in Eq. \ref{eq:timescale3}, we get
\begin{equation}\label{eq:timescalemodel1}
\ln\Big(\frac{\tau}{\tau_0}\Big)=  \frac{E}{T - T_K + \frac{Hf_0^2\tau_p}{1 + G\tau_p}},
\end{equation}
where $H = T_K \kappa_a/(\gamma\Delta C_p)$ and $G = k/\gamma$ are constants. On the other hand, one obtains $ \delta \Phi =  \Teff^{\text{sp}}/(\gamma+ k\tau_p)$ for model 2, and this gives the relaxation time as
\begin{equation}\label{eq:timescalemodel2}
\ln\Big(\frac{\tau}{\tau_0}\Big)=  \frac{E}{T - T_K + \frac{H\Teff^{sp}}{1 + G\tau_p}}.
\end{equation}
The expressions of $H$ and $G$ remain the same as earlier. $\Teff^{\text{sp}}$ is analogous to $f_0^2$.  The strength of the noise in model 2 changes with $\tau_p$, leading to opposing behaviors within the two models as a function of $\tau_p$. For example, $\tau$ and fragility decrease as $\tau_p$ increases in model 1 (Fig. \ref{theorymctrfot} c and d), whereas they increase as $\tau_p$ increases in model 2. The theoretical results agree remarkably well with simulation data for both models when $\tau_p$ is small, where an effective FDT is valid \cite{Fodor2016,parisi2005}. The theory helped rationalize some contrasting results \cite{Flenner2016,Mandal2016b} in the active glass literature. This work also highlights that the precise nature of activity is crucial. The theory has recently been extended \cite{Mandal2022a} to higher activity regime. Recently, Ref. \cite{Paul2021b} tested some of the crucial approximations of the original active RFOT theory \cite{Nandi2018a}. Specifically, Ref. \cite{Paul2021b} has computed $\xi$ in a model active glass-forming liquid using detailed FSS analysis of $tau$ s well as block analysis methods and tested the prediction of active RFOT theory (Eq.\ref{eq:lengthscale2}). Interestingly, they find that the exponent $\theta$ depends on the strength of activity in a systematic manner, much like the MCT exponents \cite{Kranz2010,Paul2021a}. Similarly, the exponent $\psi$ that relates $\tau$ with $\xi$ also depends on the strength of activity. However, the combination of the exponents that defines the $T$-dependence of the relaxation dynamics becomes independent of activity. This result explains why relaxation dynamics remain equilibrium-like \cite{nandi2018,Nandi2018a,Berthier2019c} despite the non-trivial dependence of activity on the dynamics.

{\bf RFOT theory for confluent systems:}
\label{PassiveRFOTconflu}
As discussed in Sec. \ref{confluentmodels}, the confluent systems and particulate models are fundamentally different. Usually, we neglect the particle shapes in most scenarios of physics problems. However, cell shapes are crucial as they determine many biologically significant properties \cite{Moseley2010,Jaiswal2006,Miroshnikova2018,Nguyen2022,Minc2011,Chen1997}. Some of us have recently shown that we can statistically describe the cell shape variability in a confluent monolayer \cite{Sadhukhan2022}. Moreover, the dynamics of the monolayer also depends on the cell shape variability. Many experiments have explored the glassy dynamics in such systems \cite{Park2015a,Park2016b,Giavazzi2017,Giavazzi2018,Malinverno2017,Palamidessi2019,Vishwakarma2020}. Simulations of the models discussed in Sec. \ref{confluentmodels} have also provided crucial insights \cite{Bi2014,Bi2016,Schotz2013,Sussman2018a,Paoluzzi2021,Li2021}. However, analytical theories for such systems are rare. In 2021, some of us phenomenologically extended RFOT theory to understand the glassy dynamics in confluent cellular monolayers \cite{Sadhukhan2021a}. One fundamental parameter in these systems is the target perimeter $p_0$, representing the inter-cellular interaction potential (Eq. \ref{energyfunction2}). Since interactions determine both the surface energy and the configurational entropy \cite{Parisi2010}, we can express them in terms of $p_0$ by expanding the interaction potential around a chosen $p_0$.

Reference \cite{Sadhukhan2021a} showed that the dynamics can have two distinct regimes: the low-$p_0$ regime, where the dynamics depends on $p_0$, and the large-$p_0$ regime, where the dynamics is independent of $p_0$.

 Note that an object with a given area has a minimum perimeter, $p_\text{min}$. When there is no restriction on the shape, this $p_\text{min}$ is $2\sqrt{\pi}$ corresponding to a circle of unit area. However, there is a constraint on shape for confluent systems as circles cannot fill space. The space-filling regular shape in two-dimension is a hexagon; $p_\text{min}$ is 3.722 for a hexagon with unit area.   On the other hand, there is no restriction on the maximum value of the perimeter. For a system with irregular objects, $p_\text{min}$ is higher and depends on the degree of irregularity. The low-$p_0$ regime corresponds to when $p_0 < p_\text{min}$, and the large-$p_0$ regime corresponds to when $p_0>p_\text{min}$. In the low-$p_0$ regime, cells cannot satisfy the perimeter constraint in Eq. (\ref{energyfunction2}), and the dynamics depends on $p_0$. Expanding the potential around a reference $p_0$ value, $p_0^{\text{ref}}$, and simplifying, we obtain for the low-$p_0$ regime,
\begin{equation}\label{rfotlowP0}
\ln\left(\frac{\tau}{\tau_0}\right)=\frac{k_1-k_2(p_0-p_0^{ref})}{T-T_K+\varkappa_c(p_0-p_0^{\text{ref}})}
\end{equation}
where $k_1$, $k_2$ and $\varkappa_c$ are constants \cite{Sadhukhan2021a}. Various constants in Eq. (\ref{rfotlowP0}) can be obtained by fitting the analytical form with one set of data. Once these constants are determined, one can compare the theory with simulation results. The theory agrees well with simulation data of confluent systems. As discussed earlier, one of the striking features of the confluent systems is the readily-found sub-Arrhenius relaxations \cite{Sussman2018b,Li2021}. This simple extension of the RFOT theory can also capture this behavior. One of the novel predictions of the theory is the super-Arrhenius relaxation at very low $p_0$; this is also consistent with simulations \cite{Sadhukhan2021a,Li2021}. The distinctive potential governed by the perimeter constraint, the second term in Eq. (\ref{energyfunction2}), is essential for the sub-Arrhenius behavior.

On the other hand,  if $p_0$ is large, the cells can satisfy the perimeter constraint and the second term in Eq. (\ref{energyfunction2}) becomes zero. Therefore, we expect the dynamics should be independent of $p_0$. Via a straightforward calculation, Ref. \cite{Sadhukhan2021a} obtained in this regime,
\begin{equation}\label{rfotlargeP0}
\ln\left(\frac{\tau}{\tau_0(P_0)}\right)=\frac{\Xi}{T-T_K}.
\end{equation}
Note that the high-$T$ dynamics is still $p_0$-dependent, i.e., $\tau_0$ will depend on $p_0$. But, the glassy aspects are independent of $p_0$. The theory agrees with simulation data in this regime as well.

The above theory does not contain cellular motility. However, motility is crucial in many systems. For example, the over-expressing of various oncogenes can fluidize a confluent cell monolayer from a solid-like state \cite{Malinverno2017}. This has direct consequences to health and diseases. In a recent work \cite{Sadhukhan2023}, some of us have included self-propulsion with the RFOT theory framework and investigated the effects of motility on the glassy dynamics in confluent systems.

\section{Conclusions and future perspectives}
\label{conclusion}
Active glasses have immensely enriched the field of glassy dynamics. The fact that a seemingly similar mechanism is relevant in the progression of cancer \cite{Park2015a,Atia2018,Roshal2022,Malmi-Kakkada2018}, healing of wounds \cite{Vishwakarma2020,Garcia2015}, development of embryos \cite{Mongera2018,Rode2019}, transport in cell cytoplasm \cite{Parry2014,Nishizawa2017}, and movement of molecules in dense aggregates \cite{Berthier2011a} is fascinating and surreal \cite{Park2016b,Atia2021,Berthier2019c}. These observations have motivated scientists from diverse fields to think about glassy dynamics. It extends the scope and extent of the equilibrium problem. However, there are also challenges. A quantitative and coherent understanding demands theoretical progress. Compared with the usual equilibrium particulate models of physics, these systems are immensely complex. The term `activity' has many forms: self-propulsion, confluency, change of conformation, division, apoptosis, modulation of interaction, differentiation, attachment-detachment kinetics, etc. Each of these processes is a biological marvel. But, for theoretical progress, we must learn how to formulate simple rules for mathematical description of these processes and eventually develop an analytical theory. 
The field of biological physics has shown that such exercises, though not straightforward, are possible \cite{samreview,jacques2015,Marchetti2013}.

Active systems are, by definition, out of equilibrium. For such systems, it is unclear if the well-known tools of equilibrium statistical physics are still applicable. The research of the last decades has shown that the scenario is not entirely hopeless. In the regime of low activity, generalized fluctuation-dissipation-like relations remain valid \cite{parisi2005,jacques2015,Fodor2016,petrelli2020}, and many aspects of the equilibrium glassy dynamics survive \cite{Berthier2013,Berthier2019c,Nandi2017b}. However, one must exercise caution when outside the comfort zone of equilibrium \cite{bialke2015,caprini2020,Cates2015b}. As shown earlier, while obtaining MCT for the same system via different approaches, the final form of the theory varies \cite{liluashvili2017,feng2017,Szamel2015b,Szamel2016,Flenner2016,Nandi2017b,szamel2019}. This variation is possibly due to the complex nature of the systems where the slight differences in the approximations in various approaches are significant, even though all the variants seem to agree reasonably well with simulations. A detailed comparison of these theories and finding the reasons behind the differences can bring deeper insights about the theory itself. On the other hand, the final analytical forms of the RFOT theory, obtained in a regime of linear response, are simple, although several assumptions of the theory remain unclear. Understanding these assumptions for active systems will be crucial for further understanding.

Several features make active systems qualitatively different from equilibrium systems: the long-range velocity correlations \cite{Marchetti2013,Szamel2016,caprini2020b}, giant number fluctuations \cite{Narayan2007,Giavazzi2017}, ordering transition (flocking) in spatial dimension two \cite{vicsek1995,Ramaswamy2010}, motility-induced phase separation \cite{caprini2020,Cates2013a,Cates2015b}, etc. The long-range velocity correlation survives in the dense regime. However, numerical measurements show the $T$-dependence of this correlation is relatively weak (compared to the relaxation time); this suggests that the velocity correlation remains unrelated to the glassy aspects \cite{Flenner2016,Berthier2019c}. Although giant number fluctuation shows up in confluent systems \cite{Giavazzi2017}, it is unclear if it can survive in glasses. On the other hand, flocking and phase separation are avoided in glass-forming systems. Nevertheless, the vestige of these processes can still significantly affect the glassy dynamics.

Although active systems are more complex than equilibrium systems, we can still use activity as a probe to gain crucial insights into the equilibrium problem. In this context, we discuss the specific aspect of dynamical heterogeneity (DH). Despite decades of research, a quantitative understanding of DH remains elusive. MCT predicts a divergence of the DH length scale, $\xi_D$. However, in simulation or experiments of passive glassy systems, $\xi_D$ increases by a mere factor of $5$ or so. Tests of the critical properties, where the predictions are applicable when $\xi_D\to\infty$, with such a tiny increase, in reality, is hard. By contrast, active systems in the presence of self-propulsion can show massive growth in $\xi_D$ \cite{Paul2021a}; thus, it is easier to test theoretical predictions. Moreover, the self-propelled systems are amenable to detailed theoretical treatments  \cite{Berthier2013,Nandi2017b,Paul2021a} with nonequilibrium formalism. Therefore, these systems can bring critical insights into the theories of glassy dynamics in general. 

The theoretical works for active glasses to date are mainly focused to particulate systems. However, many biologically significant processes where glassiness is vital occur in systems of cellular aggregates. For such systems, the shape of the particles is crucial \cite{Minc2011,Chen1997,Nguyen2022,Atia2018,Sadhukhan2022,Moseley2010}. Moreover, many of them are also confluent, i.e., there is no inter-particle gap in the system \cite{Farhadifar2007}. The constraint of confluency is a challenging mathematical problem \cite{weaire1986}; thus, developing theories for such systems is demanding. Most insights about these systems come from simulations of model systems \cite{Bi2016,Farhadifar2007,Fletcher2014,Sadhukhan2021a}; only some phenomenological extensions of RFOT theory exist to understand the effects of the control parameters \cite{Sadhukhan2021a}. Analytical frameworks, including some aspects of cellular shape, will be influential and valuable.

In conclusion, the dynamics in many biological systems, at varying length scales, show glassy behavior. Characterizing a glassy system is non-trivial: several characteristics must exist \cite{Berthier2011a}. The systematic exploration of glassy dynamics in biological systems dates back around two decades \cite{fabry2001} when several of the primary glassy characteristics just started to be revealed \cite{franz2000,Chaudhuri2007,IMCT}. Theoretical development in this direction is much more recent, about a decade old \cite{Berthier2013,Szamel2015b,Bi2014,Park2015a}. Note that the equilibrium problem of glassy dynamics remains unsolved, and the field continues to evolve. Active glasses enrich this field with fascinating systems, new control parameters, and different levels of complexity. Theoretical understanding of these systems becomes even more challenging. However, theories can add value in revealing patterns, trends, and new phenomena. Physics mainly concerns finding the general, universal properties of various systems. Finding the generic principles within the world of biological complexity is not straightforward, but worth pursuing, as ``{\it life out of equilibrium is typically richer than in equilibrium}" \cite{galliano2023}. Crucially, a quantitative understanding of the dynamics of these systems has far-reaching impacts and consequences.

\section{Acknowledgments}
We thank Ludovic Berthier, Satyam Pandey, and Puneet Pareek for comments on the manuscript. We also thank Puneet Pareek for Fig. 4(a). We acknowledge the support of the Department of Atomic Energy, Government of India, under Project Identification No. RTI 4007 and the computational facility of TIFR Hyderabad. SK would like to acknowledge support through the Swarna Jayanti Fellowship Grant No. DST/SJF/PSA-01/2018-19 and SB/SFJ/2019-20/05 and the MATRICS grant MTR/2023/000079 from Science and Engineering Research Board (SERB), Government of India. SKN thanks SERB for grant via SRG/2021/002014.

\bibliography{activeglassref}

\end{document}